\documentclass[journal]{IEEEtran}
\usepackage{amsfonts}
\usepackage{amssymb}
\usepackage{amsmath}
\usepackage{diagbox}
\usepackage{algorithm}
\usepackage{multirow}
\usepackage{xcolor}
\usepackage{graphicx}
\usepackage{cite}
\usepackage{exscale}
\usepackage{relsize}
\usepackage{algpseudocode}
\usepackage{graphics}
\usepackage{mathrsfs}
\usepackage{siunitx}
\usepackage{threeparttable}
\usepackage{url}
\usepackage{booktabs}
\usepackage{lipsum}
\usepackage{array}

\DeclareMathOperator*{\argmin}{argmin}
\newcommand{\bm}[1]{\mbox{\boldmath{$#1$}}}

\ifCLASSOPTIONcompsoc
\usepackage[caption=false,font=normalsize,labelfont=sf,textfont=sf]{subfig}
\else
\usepackage[caption=false,font=footnotesize]{subfig}
\fi

\usepackage[T1]{fontenc}

\title{Adversarial Reconfigurable Intelligent Surface Against Physical Layer Key Generation}

\author{Zhuangkun Wei, Bin Li, Weisi Guo

\thanks{This work is supported by the Engineering and Physical Sciences Research Council [grant number:  EP/V026763/1]. \\
Zhuangkun Wei is with the School of Aerospace, Transport, and Manufacturing, Cranfield University, MK43 0AL, UK.\\
Weisi Guo is with the School of Aerospace, Transport, and Manufacturing, Cranfield University, MK43 0AL, UK, and also with the Alan Turing Institute, London, NW1 2DB, UK.\\
Bin Li is with the Department of Information Engineering, Beijing University of Posts and Telecommunications, Beijing, 100876, China.}}

\begin{document}

\maketitle

\begin{abstract}
The development of reconfigurable intelligent surfaces (RIS) has recently advanced the research of physical layer security (PLS). Beneficial impacts of RIS include but are not limited to offering a new degree-of-freedom (DoF) for key-less PLS optimization, and increasing channel randomness for physical layer secret key generation (PL-SKG). However, there is a lack of research studying how adversarial RIS can be used to attack and obtain legitimate secret keys generated by PL-SKG. 
In this work, we show an Eve-controlled adversarial RIS (Eve-RIS), by inserting into the legitimate channel a random and reciprocal channel, can partially reconstruct the secret keys from the legitimate PL-SKG process.
To operationalize this concept, we design Eve-RIS schemes against two PL-SKG techniques used: (i) the CSI-based PL-SKG, and (ii) the two-way cross multiplication based PL-SKG. The channel probing at Eve-RIS is realized by compressed sensing designs with a small number of radio-frequency (RF) chains. Then, the optimal RIS phase is obtained by maximizing the Eve-RIS inserted deceiving channel.
Our analysis and results show that even with a passive RIS, our proposed Eve-RIS can achieve a high key match rate with legitimate users, and is resistant to most of the current defensive approaches. This means the novel Eve-RIS provides a new eavesdropping threat on PL-SKG, which can spur new research areas to counter adversarial RIS attacks.
\end{abstract}


\section{Introduction}
Wireless communications are vulnerable to diverse attack vectors due to their broadcasting nature. Traditional cryptography techniques require high computational complexity and delays to ensure confidentiality, which makes them less attractive in real-time and lightweight systems \cite{8883127}. To secure the wireless channels, a variety of physical layer security (PLS) techniques have been proposed and widely studied in the last decade. 

\subsection{Literature Review}
PLS techniques can be categorized as key-less PLS and physical layer secret key generation (PL-SKG).

\subsubsection{Key-Less PLS}
Key-less PLS tries to maintain the superiority of legitimate channels by maximizing the secrecy rate (via e.g., the beamforming vector \cite{9520776}, the trajectory of autonomous systems \cite{9656117}, the anti-jamming artificial noise \cite{8456560}, the spin modulation, etc.). The challenge lies in the high dependency on additional positioning data and the lack of guarantee of a feasible solution, especially when combined with real-world constraints. 

\subsubsection{PL-SKG}
Another family is PL-SKG, which leverages the reciprocal channel randomness to generate shared secret keys \cite{6739367,7393435,4036441,5422766,5371757}. Most of the PL-SKG schemes exploit the channel state information (CSI) as the common random feature, e.g., the received signal strength (RSS) \cite{7393435}, the channel phases \cite{7933224}, and the channel frequency response \cite{8293762}. In these cases, two legitimate nodes (e.g., Alice and Bob) are required to send public pilot sequences to each other and pursue channel estimations to acquire these common CSI, which will then be passed to the quantization \cite{6171198,mathur2008radio}, information reconciliation \cite{10.1007/3-540-48285-7_35} and privacy amplification \cite{impagliazzo1989pseudo} modules for key generation. 

One challenge on PL-SKG is that the secret key rate cannot meet the industrial requirement due to insufficient channel randomness (e.g., RSS variations and small-scale channel scattering \cite{9000831}), \textcolor{blue}{although optimization algorithms (e.g., power allocation \cite{lu2022secret}) can be used to improve the legitimate SKR}. To address this, one-way based PL-SKG has been proposed by the works in \cite{8254029,lou2017secret,7582525}, whereby one legitimate node (e.g., Alice) sends public pilots and Bob sends random signals. In this way, the common feature is Alice's received signals, which, at Bob's end, can be constructed by his channel estimation result and his sending random signals. As such, the feature randomness not only involves the random CSI but is enhanced by Bob's transmitted random signal, and thereby improving the SKR. 

Inspired by the one-way randomness enhancement, the works in \cite{7593219,8057119,8424614,7794581} further promote the SKR by leveraging the two-way random signals, whereby Alice and Bob send random pilots to each other and cross multiply their sent and received signals as the common feature (known as two-way cross multiplication method). In this view, the randomness of the common feature is further enhanced by two random spaces, and therefore leads to a higher SKR as opposed to one-way based and CSI-based PL-SKGs. Despite these advances, \textcolor{blue}{the improved SKR schemes are still not enough to approach the current Gbps levels of the transmission rate (i.e., making one information bit have one unique secret key for encryption), which renders as the main challenge to impede PL-SKG from civilian and commercial use.}

\subsubsection{When PLS meets RIS}
Reconfigurable intelligent surface (RIS) has been recently proposed to change and adjust the communication channels to improve the communication quality of services (QoS) \cite{9326394,8910627,8811733,ma2019controllable}. In the context of PLS, RIS can (i) \textcolor{blue}{serve as a new degree-of-freedom (DoF)} for optimizing the secrecy rate in key-less PLS \cite{9520295,9201173}, and (ii) increase channel randomness by its phase controller for secret key generation \cite{9442833,9361290,9360860,staat2020intelligent}. 
To be specific, by randomly assigning the RIS phase in each channel estimation round, the reciprocal randomness of legitimate channels can be artificially enhanced, enabling a fast generation of the shared secret key. Based on this idea, \cite{9442833} computes the SKR of RIS-secured low-entropy channel, and \cite{9298937} further designs an optimal RIS phase set by maximizing the theoretical SKR. 

The advance of RIS also provides new attack and eavesdropping potentials. This can be categorized as attackers that (i) destroy or (ii) maintain the channel reciprocity, where the former aims to ruin the legitimate PL-SKG, and the latter tries to obtain the legitimate secret keys. 
\textcolor{blue}{For example, the attacker in \cite{9625442} controlled a RIS to damage PL-SKG at legitimate parties, by destroying the channel reciprocity via a fast change of RIS phase in the legitimate key generation process. However, the attackers that destroy the channel reciprocity cannot obtain legitimate secret keys without being detected.}
In this work, we focus on the second category, i.e., how a RIS can reconstruct the legitimate secret keys, by generating and inserting a deceivingly reciprocal and random channel into legitimate channels (named as Eve-RIS). 
\textcolor{blue}{Note that a similar idea, known as the secret key leakage attack by an adversarial RIS, has been proposed in the work \cite{9771319}. However, the critical detailed implementation (e.g., how the RIS pursues channel estimation and how to optimize the deceiving channel) is missing. Besides, the countermeasure claimed in \cite{9771319}, (i.e., the two-way cross multiplication method in \cite{9000831}) is actually what we are going to attack in this work (in Section III. B).}

\textcolor{blue}{Indeed, an untrust relay can pursue a similar man-in-the-middle (MiM) attack by generating and inserting a reciprocal channel to obtain legitimate secret keys. However, such a threat can be easily addressed by designing appropriate relay pilot transmission protocols \cite{thai2015physical,letafati2020new}. 
RIS, given its reflective property (unable to actively send pilots for protocol and authentication purposes), is naturally resistant to these countermeasures, and therefore paves a way to realize this MiM attack in a more concealed way. Also, given its ability to manipulate channels, we show (in Fig. \ref{figs2_3}) that a passive adversarial RIS (with $1600$ elements) can achieve a comparable eavesdropping performance with the untrust relay using $60$dB amplifying gain. In this view, our proposed Eve-RIS provides a novel instance of MiM attacks, which is more energy-efficient and in a more concealed manner. A comprehensive comparison is provided in Section IV. A.}

\subsection{Contributions \& Paper Structure}
In this work, we aim to design an adversarial Eve-controlled RIS-based eavesdropping scheme. Eve-RIS aims to generate and insert a deceivingly random and reciprocal channel between Alice and Bob, so that their CSI-based secret key will be partially inferable to Eve. The main novel contributions are listed in the following.  

(1) We show that an Eve-controlled RIS, by inserting a deceivingly random and reciprocal channel into legitimate channels, can partially reconstruct the PL-SKG based secret keys from legitimate users. The resulting theoretical key match rate between the proposed Eve-RIS and the legitimate users is deduced, whose geometry and qualitative properties provide insights for further designs and implementations. 

(2) Operationalizing this, we design two eavesdropping schemes against (i) the CSI-based PL-SKG, and (ii) the two-way cross multiplication-based PL-SKG, respectively. Equipped with a small number of radio-frequency (RF) chains, compressed sensing-based baseband channel probing and feature extraction methods are designed for Eve-RIS. Then, the optimal RIS phase is obtained by maximizing the Eve-RIS inserted deceiving channel.

(3) We perform a comprehensive comparison between our designed Eve-RIS and other popular attackers, which are categorized by their ability to maintain or destroy the channel reciprocity. The main difference lies in (i) their inability to manipulate their CSI-based attack and thereby suffering from the severe cascaded channel attenuation, and (ii) the unsuitability of their defensive approaches to the proposed Eve-RIS in this work

(4) We evaluate our proposed Eve-RIS via simulations. The results show that even with a passive RIS, our proposed Eve-RIS can achieve a high key match rate with legitimate users, and is resistant to most of the current defensive approaches. \textcolor{blue}{As such, our proposed Eve-RIS provides a new eavesdropping threat on PL-SKG, which should be seriously considered by further secret key designs to protect the confidentiality of wireless communications.}

The rest of this work is structured as follows. In Section II, we describe the Eve-RIS model in the Alice-Bob scenario. In Section III, we elaborate on our design of Eve-RIS schemes against channel estimation based and two-way cross multiplication based PL-SKGs. In Section IV, we compare our designed Eve-RIS with other popular attackers, from the conceptual perspective. In Section V, we show our simulation results. We finally conclude this work in Section VI. 

In this work, we use bold lower-case letters for vectors, and bold capital letters for matrices. We use $\|\cdot\|_2$ to denote the $2$-norm, $\|\cdot\|_0$ to denote the $0$-norm, and $diag(\cdot)$ to diagonalize a vector. We denote $mod(\cdot,\cdot)$ as the modulus operator and $\lfloor\cdot\rfloor$ is to truncate the argument. The matrix transpose, conjugate transpose, element-wise conjugate, and Hadamard product operators are denoted as $(\cdot)^T$, $(\cdot)^H$, $(\cdot)^*$, and $\odot$. $\mathbb{E}(\cdot)$ and $\mathbb{D}(\cdot)$ represent the expectation and variance. $\mathcal{CN}(\mu,2\sigma^2)$ is to represent the complex Gaussian distribution with mean as $\mu$ and variance as $2\sigma^2$.

\begin{figure*}[!t]
\centering
\includegraphics[width=7in]{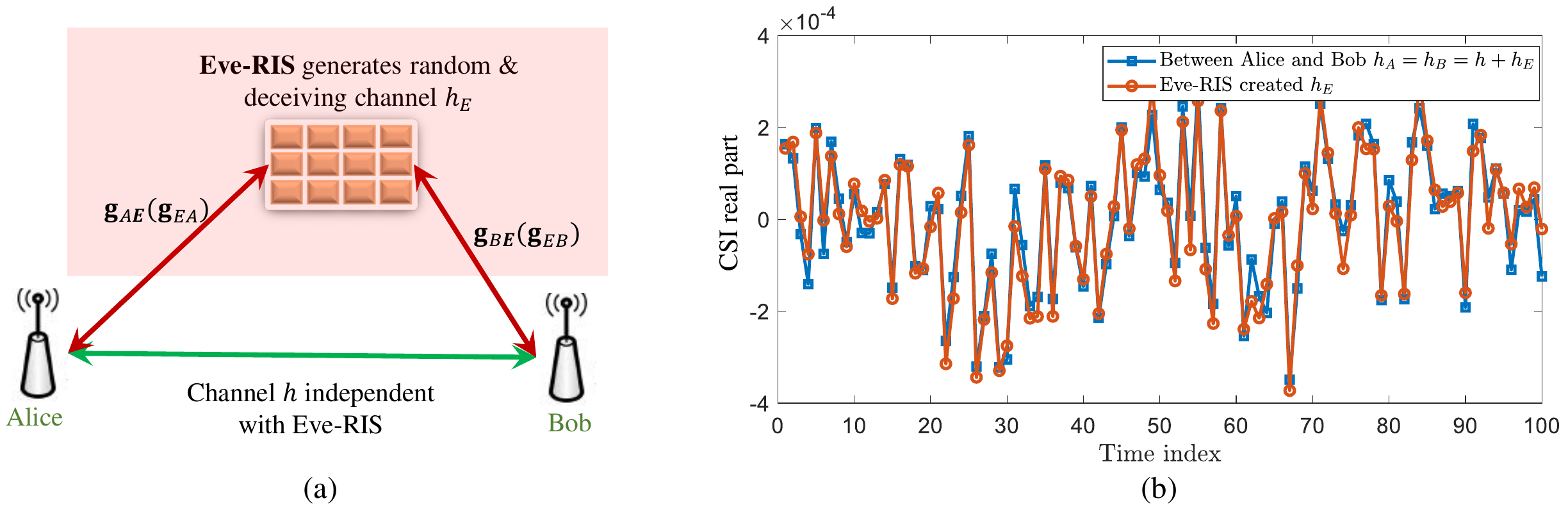}
\caption{Sketch of Eve-RIS: (a) the deployment of Eve-RIS to generate and insert a deceiving and random channel $h_E$, serving as a part of the legitimate channel between Alice and Bob, i.e., $h_A=h_B=h+h_E$, (b) illustration of $h_A=h_B\approx h_E$ by their real parts.}
\label{fig1}
\end{figure*}

\section{System Model \& Problem Formulation}

In this work, two legitimate users Alice and Bob are to generate a shared secret key, leveraging the reciprocal channels between them. Eve pursues eavesdropping by generating and inserting a random channel between Alice and Bob, which is achieved by a RIS, \textcolor{blue}{with uniform planar array (UPA)} of size $M=M_x\times M_y$ (see Fig. \ref{fig1}(a)). \textcolor{blue}{Different from the general RIS design in \cite{8910627}, we refer to the hardware architecture in \cite{taha2021enabling} to facilitate RIS's baseband channel estimation: A few of reflective elements are deployed with channel sensors, each of which connects to an RF chain for baseband measurements and signal processing.}

The direct channels are modeled via a Rician fading model. We express the direct channel between Alice and Bob (irrelevant with Eve-RIS) as \cite{6847111,9034493}:
\begin{equation}
\label{eq01}
    h\sim\mathcal{CN}(0,2\sigma_h^2),~2\sigma_h^2=C_0\cdot d_{AB}^{-\alpha_{N}}.
\end{equation}
In Eq. (\ref{eq01}), $C_0$ is the path loss at the reference distance (i.e., $1m$), $d_{AB}$ is the LoS distance between Alice and Bob, and $\alpha_{N}$ is the NLoS path-loss exponent. Here, the static LoS component is removed, as it is trivial for PL-SKG. 
{\color{blue}The channels from Alice and Bob to Eve-RIS are expressed as $\mathbf{g}_{aE}\sim\mathcal{CN}(\mathbf{g}_{aE}^\text{(LoS)},2\bm{\Sigma}_{aE})~a\in\{A,B\}$, which are modeled as \cite{9300189}:
\begin{equation}
\label{eq02}
\begin{aligned}
&\mathbf{g}_{aE}=\mathbf{g}_{aE}^\text{(LoS)}+\sum_{n=1}^{\iota}\frac{\rho_{aE,n}}{\sqrt{\iota}}\cdot\mathbf{u}\left(el_{aE,n},az_{aE,n}\right)\\
    &\mathbf{g}_{aE}^\text{(LoS)}=
    \sqrt{C_0\cdot d_{aE}^{-\alpha_L}}\cdot\mathbf{u}\left(el_{aE}^{(LoS)},az_{aE}^{(LoS)}\right)\\
    &\rho_{aE,n}\sim\mathcal{CN}(0,C_0\cdot d_{aE}^{-\alpha_N})
\end{aligned}    
\end{equation}
In Eq. (\ref{eq02}), $d_{aE}$ is the LoS distance between $a$ to Eve-RIS, and $\alpha_L$ and $\alpha_N$ denote the LoS and NLoS path-loss exponents, respectively. $\iota$ is the number of NLoS Rayleigh paths and $\rho_{aE,n}$ is the gain for $n$th path. 
$\mathbf{u}(el,az)\triangleq[\exp(j\mathbf{a}(el,az)\mathbf{l}_1),\cdots,\exp(j\mathbf{a}(el,az)\mathbf{l}_M)]^T$, with $\mathbf{a}(el,az)\triangleq\frac{2\pi}{\lambda}[\sin(el)\cos(az),\sin(el)\sin(az),\cos(el)]$ and $\mathbf{l}_m\triangleq[0,mod(m-1, M_x)d,\lfloor (m-1)/M_y\rfloor d]^T$. $el_{aE,n},az_{aE,n}\in[-\pi/2,\pi/2]$ are the half-space elevation and azimuth angles of $n$th path. For the structure of RIS, a square shape element is used with the size as $d\times d$, where $d=\lambda/8$ is set (i.e., less than half-wavelength $\lambda/2$ \cite{9300189,tsilipakos2020toward,8936989}).}

With the modeling of the direct channels, the Eve-RIS generated channel, denoted as $h_E$, and its combined channels from Bob to Alice (Alice to Bob), denoted as $h_A$ ($h_B$), can be expressed as follows:
\begin{equation}
\label{eq1}
\begin{aligned}
    &h_E=\mathbf{g}_{BE}^T\cdot diag(\mathbf{w})\cdot\mathbf{g}_{AE}\sim\mathcal{CN}(0,2\sigma_E^2),\\
    &h_A=h_B=h+h_E.
\end{aligned}
\end{equation}
In Eq. (\ref{eq1}), $\mathbf{w}=\sqrt{A_E}[\exp(j\theta_1),\cdots,\exp(j\theta_M)]^T$ is the phase vector of the Eve-RIS, where $A_E\in\mathbb{R}^+$ is the amplifying gain, and $\theta_m\sim\mathcal{U}[0,2\pi)$ with $m\in\{1,\cdots,M\}$ is the random phases. As such, the probability density distribution (PDF) of $h_E$ can be approximated as:
\begin{equation}
\label{eq3}
    h_E\sim\mathcal{CN}(\mu_E,2\sigma_E^2),
\end{equation}
for a large number of Eve-RIS elements (see Appendix A for details).  In Eq. (\ref{eq3}), $\mu_E$ and $\sigma_E^2$ are computed as:
\begin{equation}
\label{eq3_4}
\mu_E=
\begin{cases}
0, &\text{random } \mathbf{w},\\
\left(\mathbf{g}_{BE}^{\text{(LoS)}}\right)^T\cdot diag(\mathbf{w})\cdot\mathbf{g}_{AE}^{\text{(LoS)}},&\text{fixed } \mathbf{w},
\end{cases}
\end{equation}
{\color{blue}
\begin{equation}
\label{eq4}
    \sigma_E^2=
    \begin{cases}
    0.5A_E\cdot M\cdot C_0^2\cdot d_{AE}^{-\alpha_L}\cdot d_{BE}^{-\alpha_L}, &\text{random } \mathbf{w},\\
    \mathbf{w}^H\cdot\mathbf{G}\cdot\mathbf{w}, & \text{fixed } \mathbf{w}
    \end{cases}
\end{equation}
where $\mathbf{G}\triangleq2\bm{\Sigma}_{AR}\odot\bm{\Sigma}_{BR}+ diag(\mathbf{g}_{BR}^\text{(LoS)})^*\bm{\Sigma}_{AR}diag(\mathbf{g}_{BR}^\text{(LoS)})+diag(\mathbf{g}_{AR}^\text{(LoS)})^*\bm{\Sigma}_{BR}diag(\mathbf{g}_{AR}^\text{(LoS)})$. The detailed deductions are provided in Appendix \ref{appendix1}. From Eq. (\ref{eq4}), the variance $\sigma^2_E$ is determined by (i) the multiplication of variances of two sub-channels, i.e., $C_0d_{AE}^{-\alpha_L}\cdot C_0d_{BE}^{-\alpha_L}$, known as the cascaded channel attenuation, (ii) the number of RIS element $M$, and (iii) the RIS programmable phase vector $\mathbf{w}$. We will further show how these will be used to design our Eve-RIS schemes.}

With the formulated model, the purpose of this work is to design how the Eve-RIS can eavesdropping the secret key between Alice and Bob. We will study two key generation cases: (i) SKG using channel estimation results, and (ii) SKG using the two-way method.

\section{Designs of Eve-RIS against PL-SKG}

We first give a sketch of the Eve-RIS scheme. As Alice and Bob use their reciprocal channels, i.e., $h_A=h_B$, for secret key generation, they do not know that $h_A$ and $h_B$ contains the Eve-RIS's deceiving channel $h_E$, i.e., $h_A=h_B=h+h_E$ (see Fig. \ref{fig1}(b)). \textcolor{blue}{In this view, a large variance of $h_E$, i.e., $\sigma_E^2$, will lead to a high correlation coefficient of $h_E$ and $h_A$ ($h_B$), i.e., $corr(h_E,h_A)=\sigma_E^2/\sqrt{\sigma_E^2+\sigma_h^2}$, and therefore a high secret key match rate between Eve-RIS and Alice (Bob).} 

To be specific, we consider a general two-threshold quantization method, i.e., \cite{mathur2008radio}
\begin{equation}
\label{key}
    k_a=\begin{cases}
    1,~z_a>\gamma_1^{(a)},\\
    0,~z_b<\gamma_0^{(a)},
    \end{cases} a\in\{A, B, E\},
\end{equation}
where $\gamma_1^{(a)}=\mathbb{E}(z_a)+\beta\sqrt{\mathbb{D}(z_a)}$ and $\gamma_0^{(a)}=\mathbb{E}(z_a)-\beta\sqrt{\mathbb{D}(z_a)}$ are the upper and the lower quantization thresholds, with quantization threshold parameter $\beta\in[0,0.5)$. In Eq. (\ref{key}), $z_a$ can be either $Re[h_a]$ or $Im[h_a]$, or the combination of $Re[h_a]$ and $Im[h_a]$. To simplify the further analysis, we assign $z_a=Re[h_a]$. {\color{blue}The theoretical key match rate between Alice and Eve can be computed as:
\begin{equation}
\label{theo1_1}
\begin{aligned}
    &Pr\{k_A=k_E\}\\
    =&\sqrt{\frac{2}{\pi}}\int_{\beta}^{+\infty}\Phi\left(-\beta\sqrt{\frac{\sigma_E^2}{\sigma_h^2}+1}+\frac{\sigma_E}{\sigma_h}\zeta\right)\exp\left(-\frac{\zeta^2}{2}\right)d\zeta,
\end{aligned}
\end{equation}
where $\Phi(\cdot)$ is the cumulative distribution function of a normal distribution (see Appendix \ref{append2} for detailed deduction).
It is noteworthy that a more compact version of Eq. (\ref{theo1_1}) is not available, but it can still provide insights for the design and further implementation of Eve-RIS. One illustration of Eq. (\ref{theo1_1}) is provided in Fig. \ref{fig1_2}.

First, the key match rate $\Pr\{k_A=k_E\}$ is monotonically increasing with the ratio between the variances of the Eve-RIS generated channel and the direct legitimate channel, i.e., $\sigma_E^2/\sigma_h^2$ (proved in Appendix \ref{append2} and shown in Fig. \ref{fig1_2}). Specially, we have, 
\begin{equation}
\lim_{\sigma_E^2/\sigma_h^2\rightarrow+\infty}Pr\{k_A=k_E\}=Pr\{k_A=k_B\}=2\Phi(-\beta),
\end{equation}
suggesting that in the wave-blockage scenario, the key match rate between Eve-RIS and legitimate users reaches the limit, i.e., $Pr\{k_A=k_E\}=Pr\{k_A=k_B\}$, which is determined by the quantization parameter $\beta$ set by the legitimate users. 

Second, $\Pr\{k_A=k_E\}$ at first increases faster and then gradually, with the increase of $\sigma_E^2/\sigma_h^2$ (shown in Fig. \ref{fig1_2}). This can be proved by the fact that its second-order derivative is less than $0$ (see Appendix \ref{append2} for details). Such a phenomenon indicates the potential of the Eve-RIS, whereby a slight manipulation of its deceiving channel can make a huge legitimate secret key leakage.

\begin{figure}[!t]
\centering
\includegraphics[width=3.5in]{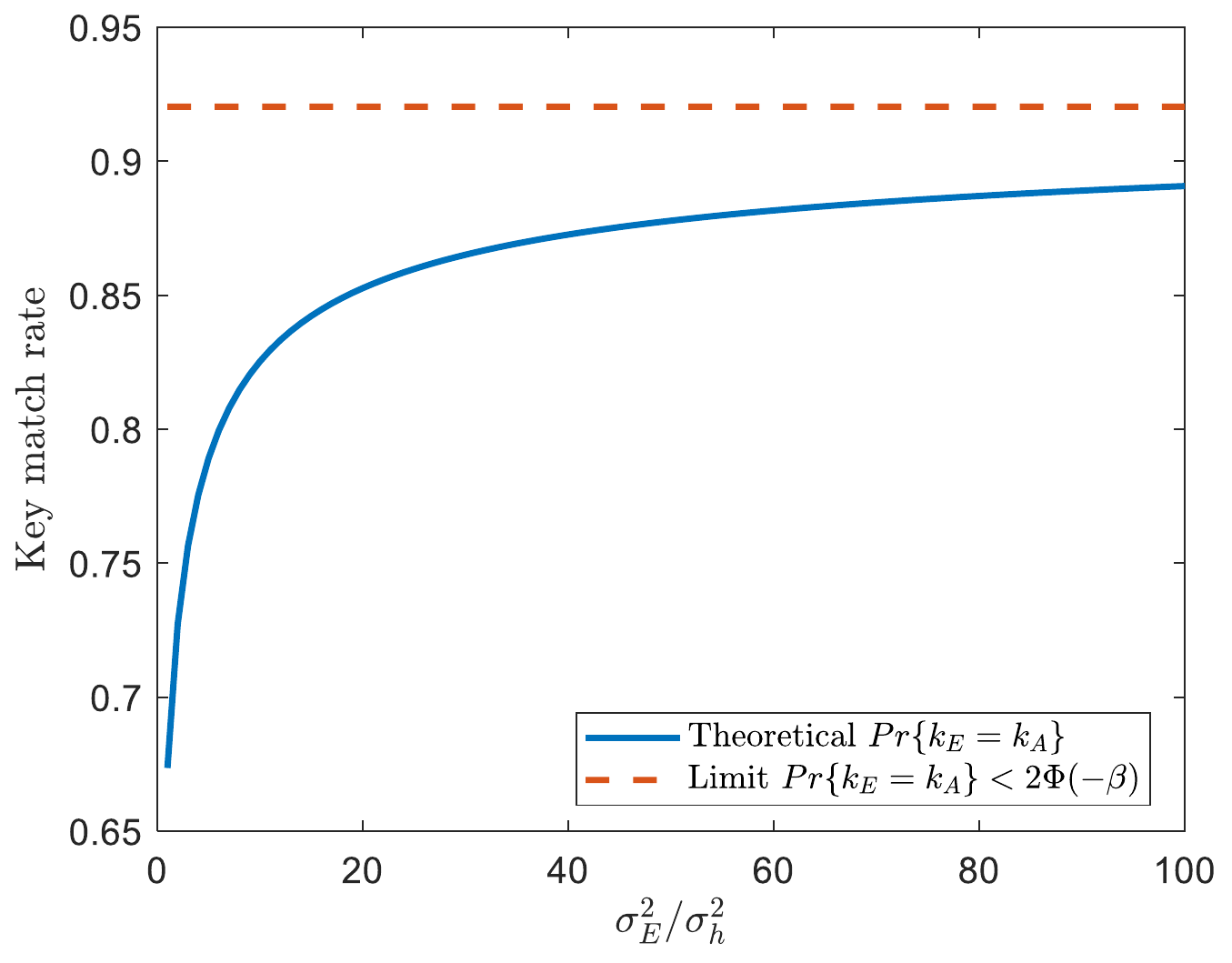}
\caption{Theoretical key match rate between proposed Eve-RIS and legitimate users, i.e., $Pr\{k_A=k_E\}$, which serves as a reference for the implementation of Eve-RIS to determine the variance of its generated deceiving channel, i.e., $\sigma_E^2$.  }
\label{fig1_2}
\end{figure}

Third, from the implementation view, Eq. (\ref{theo1_1}) provides a theoretical mapping between the Eve-RIS's achievable key match rate and its controlling variable, i.e., $\sigma_E^2$. Such a mapping serves as the reference for the design of Eve-RIS to determine an appropriate $\sigma_E^2$, given the specific requirement of the eavesdropping key match rate, i.e., $Pr\{k_A=k_E\}$. Leveraging this, further detailed implementation to optimize $\sigma_E^2$ can be pursued (e.g., balancing the number of RIS elements that will be used and the active reflecting gain, as they both increase $\sigma_E^2$ from Eq. (\ref{eq4}), and maximizing $\sigma_E^2$ by RIS phase).}

We next elaborate on the details of our Eve-RIS to attack two popular PL-SKG schemes.

\subsection{Eavesdropping CSI-based PL-SKG}
\subsubsection{PL-SKG using CSI Estimation}
We first show the process of PL-SKG using the estimations of the legitimate reciprocal channel. In this case, Alice and Bob estimate the reciprocal channel in time-division duplex (TDD) mode, whereby in each channel estimation round, the channel between Alice and Bob remains unchanged. 
In odd and even time slots, Alice and Bob respectively send pilot sequence $\mathbf{x}_A,\mathbf{x}_B\in\mathbb{C}^{1\times L}$. Then, the channels estimated at Alice and Bob, denoted as $\hat{h}_A$ and $\hat{h}_B$, are \cite{9442833,9361290,9360860,staat2020intelligent}:
\begin{equation}
\label{legitimate1}
\begin{aligned}
    \hat{h}_A&=\frac{\mathbf{y}_A\cdot\mathbf{x}_B^H}{\|\mathbf{x}_B\|_2^2}=(h+h_E)+\hat{n}_A,\\
    \hat{h}_B&=\frac{\mathbf{y}_B\cdot\mathbf{x}_A^H}{\|\mathbf{x}_A\|_2^2}=(h+h_E)+\hat{n}_B,
\end{aligned}
\end{equation}
In Eq. (\ref{legitimate1}), $\mathbf{y}_A=h_A\cdot\mathbf{x}_B+\mathbf{n}_A$ and  $\mathbf{y}_B=h_B\cdot\mathbf{x}_A+\mathbf{n}_B$ are the received signals at Alice and Bob, with $\mathbf{n}_A,\mathbf{n}_B\sim\mathcal{CN}(0,2\sigma_n^2\mathbf{I}_L)$, the receiving noise components. $\hat{n}_A\sim\mathcal{CN}(0,2\sigma_n^2/\|\mathbf{x}_B\|_2^2)$, $\hat{n}_B\sim\mathcal{CN}(0,2\sigma_n^2/\|\mathbf{x}_A\|_2^2)$ are the estimating noises. As such, leveraging the common channel estimations, i.e., $\hat{h}_A$ and $\hat{h}_B$, secret keys can be generated from Alice and Bob, by replacing $z_A$ and $z_B$ with $\hat{z}_A=Re[\hat{h}_A]$ and $\hat{z}_B=Re[\hat{h}_B]$, respectively, in Eq. (\ref{key}).  

\subsubsection{Eavesdropping design}
The purpose of Eve-RIS is to generate and insert a deceivingly random channel $h_E$ that contributes to part of $h_A$ and $h_B$. To do so, for each channel legitimate estimation round of Alice and Bob, Eve-RIS assigns a RIS phase vector $\mathbf{w}$ by either random or optimized strategies. Here the random strategy is to select $m$th elemental phase of $\mathbf{w}$ identically and randomly over $[0,2\pi)$, i.e., $\theta_m\in\mathcal{U}[0,2\pi)$. The optimized strategy will be elaborated on in Section III. C. In order to hold randomness and reciprocity, such a $\mathbf{w}$ will remain unchanged during one Alice-Bob channel estimation round but will change independently for different channel estimation rounds. 

{\color{blue}Eve-RIS will pursue estimations of Alice to Eve-RIS, and Bob to Eve-RIS channels, i.e., $\mathbf{g}_{AE}$ and $\mathbf{g}_{BE}$, by their sending pilots. According to the RIS architecture in \cite{taha2021enabling}, the elements with active channel sensors receive the transmitted signals and feed them into the RF chains for further baseband channel estimation. Then, the compressed sensing-based channel estimation is pursued, due to the sparse representation of $\mathbf{g}_{AE}$ and $\mathbf{g}_{BE}$ in the beamspace dictionary, induced by the small number of Rayleigh scattering paths, i.e., $\iota\leq5$ \cite{alexandropoulos2020hardware}. As such, the Eve-RIS RF chain received baseband signals from Alice and Bob in one channel estimation round are:
\begin{equation}
\label{cs1}
\begin{aligned}
\mathbf{Y}_E^{(A)}&=\mathbf{C}\cdot\mathbf{D}\cdot\mathbf{s}_{AE}\cdot\mathbf{x}_A+\mathbf{N}_E^{(A)},\\    \mathbf{Y}_E^{(B)}&=\mathbf{C}\cdot\mathbf{D}\cdot\mathbf{s}_{BE}\cdot\mathbf{x}_B+\mathbf{N}_E^{(B)},\\
\end{aligned}
\end{equation}
where $\mathbf{N}_E^{(A)}$ and $\mathbf{N}_E^{(B)}$ are the noise components, whose elements are i.i.d complex Gaussian distributed with variance $\sigma_n^2$. In Eq. (\ref{cs1}), $\mathbf{D}\in\mathbb{C}^{M\times D}$ is the dictionary designed by the spanning of the RIS beamspace:
\begin{equation}
\label{dictionary}
\mathbf{D}=\left[\mathbf{u}\left(el_1,az_1\right),\cdots,\mathbf{u}\left(el_D,az_D\right)\right],
\end{equation}
where the pairs $(el_1,az_1),\cdots,(el_D,az_D)$ evenly enumerate the joint space of azimuth and elevation angles, i.e., $[-\pi/2,\pi/2]\times[-\pi/2,\pi/2]$. With the design of the dictionary, we have $\mathbf{g}_{AE}=\mathbf{D}\cdot\mathbf{s}_{AE}$ and $\mathbf{g}_{BE}=\mathbf{D}\cdot\mathbf{s}_{BE}$ in Eq. (\ref{cs1}), where $\mathbf{s}_{AE},\mathbf{s}_{BE}\in\mathbb{C}^D$ are $\iota\leq5$-sparse due to the fact that the number of their Rayleigh paths is less than $5$ \cite{alexandropoulos2020hardware}. 

In Eq. (\ref{cs1})$, \mathbf{C}\in\mathbb{R}^{C\times M}$ is the sensing matrix where each row only has one nonzero element, representing one entry of RIS element equipped with channel sensor and RF chain. The selection of $\mathbf{C}$ should ensure the restricted isometry property (RIP) \cite{candes2005decoding,candes2008restricted}, i.e., the condition number of any $2\iota$ columns of $\mathbf{CD}$ should be smaller than a threshold, which is an NP-hard problem. Here, a sub-optimal greedy strategy is proposed as:
\begin{equation}
\label{selection}
\mathcal{C}\leftarrow\mathcal{C}\cup\{n\},~n=\argmin_{i}cond\left(\mathbf{D}_{\mathcal{C}+\{i\}, :}\right),
\end{equation}
where $\mathcal{C}$ is the set of the columns of the non-zero entries in $\mathbf{C}$, and $\mathbf{D}_{\mathcal{C}, :}$ is the submatrix of $\mathbf{D}$ whose rows are selected by $\mathcal{C}$. Note that the selection in Eq. (\ref{selection}), i.e., the deployment of sensors and RF chains, is an off-line procedure, since $\mathbf{D}$ is fixed as the structure of RIS is determined, and its time and complexity consumption will not affect the real-time channel estimation. The number of channel sensors and RF chains, i.e., $|\mathcal{C}|=C$, is generally set as a little larger than $2\times \iota$. In the context of channel estimation, this number can be further reduced by extending the spatial measurements via the large time-span pilots \cite{alexandropoulos2020hardware}.

From the baseband measurements in Eq. (\ref{cs1}), and the sparse representation by Eq. (\ref{dictionary}), the compressed sensing-based channel estimations at Eve-RIS are pursued by:
\begin{equation}
\label{cs2}
\begin{aligned}
    &\min_{\mathbf{s}_{AE}}\left\|\frac{\mathbf{Y}_E^{(A)}\cdot\mathbf{x}_A^H}{\|\mathbf{x}_A\|_2^2}-\mathbf{C}\mathbf{D}\cdot\mathbf{s}_{AE}\right\|_2^2,~s.t.,~\|\mathbf{s}_{AE}\|_0\leq\iota\\
   &\min_{\mathbf{s}_{BE}}\left\|\frac{\mathbf{Y}_E^{(B)}\cdot\mathbf{x}_B^H}{\|\mathbf{x}_B\|_2^2}-\mathbf{C}\mathbf{D}\cdot\mathbf{s}_{BE}\right\|_2^2,~s.t.,~\|\mathbf{s}_{BE}\|_0\leq\iota\\
\end{aligned}
\end{equation}
To solve Eq. (\ref{cs2}), one may loosen the constraints by $1$-norm and then adopt the subgradient method, or use the orthogonal matching pursuit (OMP) algorithm \cite{6006641}. 

By denoting the solutions of Eq. (\ref{cs2}) as $\hat{\mathbf{s}}_{AE}$ and $\hat{\mathbf{s}}_{BE}$, the channel estimation results of $\mathbf{g}_{AE}$ and $\mathbf{g}_{BE}$ are:
\begin{equation}
\label{est1}
\begin{aligned}
\hat{\mathbf{g}}_{AE}=&\mathbf{D}\cdot\hat{\mathbf{s}}_{AE},\\
\hat{\mathbf{g}}_{BE}=&\mathbf{D}\cdot\hat{\mathbf{s}}_{BE}.
\end{aligned}
\end{equation}}
With the help of Eq. (\ref{est1}), Eve-RIS generated channel $h_E$ can be estimated as:
\begin{equation}
\label{est3}
\begin{aligned}
    \hat{h}_E&=\hat{\mathbf{g}}_{BE}^T\cdot diag(\mathbf{w})\cdot\hat{\mathbf{g}}_{AE}. 
\end{aligned}
\end{equation}
After the estimation of the Eve-RIS generated deceiving channel, i.e., $\hat{h}_E$ in Eq. (\ref{est3}), Eve can obtain the secret key by replacing $z_E$ with $\hat{z}_E=Re[\hat{h}_E]$ of the quantization method in Eq. (\ref{key}).

\subsection{Eavesdropping PL-SKG using two-way method}\label{twoway}
\textcolor{blue}{Two-way PL-SKG leverages the random pilots sent from legitimate users to pursue channel randomization. This thereby prevents most of the untrust relays and spoofing attackers that require exact channel estimations, e.g., \cite{letafati2021deep,eberz2012practical,pan2021man,8626511}. In this part, we will show how Eve-RIS can obtain the legitimate secret keys generated by two-way PL-SKG.} 

\subsubsection{PL-SKG using two-way method}
In the two-way method, Alice and Bob send random pilots to each other in TDD mode, assigned as $q_A,q_B\in\mathbb{C}$. Here, several designs of the distribution of $q_A$ and $q_B$ have been made in the work \cite{8424614}, but in this work, the specific assignment will not affect our eavesdropping design. Then, Alice and Bob multiply their transmitted and received signals as their common features for further key quantization, i.e., \cite{8424614,8798662}
\begin{equation}
\label{twoway1}
\begin{aligned}
    \hat{\phi}_A=v_A\cdot q_A=(h+h_E)\cdot q_A\cdot q_B+\hat{\epsilon}_A\\
    \hat{\phi}_B=v_B\cdot q_B=(h+h_E)\cdot q_A\cdot q_B+\hat{\epsilon}_B,
\end{aligned}
\end{equation}
where $v_A=h_A\cdot q_B+n_A$ and $v_B=h_B\cdot q_A+n_B$ are the received signals at Alice and Bob, with $n_A,n_B\sim\mathcal{CN}(0,2\sigma_n^2)$, the received noises. $\hat{\epsilon}_A=\epsilon_A\cdot q_A$ and $\hat{\epsilon}_B=\epsilon_B\cdot q_B$ are denoted for simplification. As such, by replacing $z_A$ and $z_B$ with $\hat{\varphi}_A=Re[\hat{\phi}_A]$ and $\hat{\varphi}_B=Re[\hat{\phi}_B]$, respectively, in Eq. (\ref{key}), legitimate secret keys between Alice and Bob can be generated. 

\subsubsection{Eavesdropping design}
\textcolor{blue}{The Eve-RIS design against two-way based PL-SKG shares similar parts to that against CSI-based PL-SKG, whereby a random or deliberately optimized phase vector $\mathbf{w}$ is assigned for each two-way key generation round. The difference is how Eve-RIS reconstructs the common feature, as exact channel estimations are unavailable due to the random pilots.}

Given the equipped channel sensors and RF chains, the received baseband signals from Alice and Bob are:
\begin{equation}
\label{cs3}
\begin{aligned}
    \mathbf{r}_E^{(A)}&=\mathbf{C}\cdot\mathbf{D}\cdot\mathbf{s}_{AE}\cdot q_A+\bm{\varepsilon}_E^{(A)},\\
    \mathbf{r}_E^{(B)}&=\mathbf{C}\cdot\mathbf{D}\cdot\mathbf{s}_{BE}\cdot q_B+\bm{\varepsilon}_E^{(B)},
\end{aligned}
\end{equation}
with $\bm{\varepsilon}_E^{(A)},\bm{\varepsilon}_E^{(B)}\sim\mathcal{CN}(0,2\sigma_n^2\mathbf{I}_C)$ the noise components. From Eq. (\ref{cs3}), it is seen that an accurate estimation of $\mathbf{s}_{AE}$ ($\mathbf{s}_{BE}$) is unavailable, due to the involvement of the unknown random pilots $q_A$ ($q_B$). However, it is noticed from Eq. (\ref{twoway1}) that with a large variance of Eve-RIS's deceiving channel, i.e., $\sigma_E^2$, the following approximation holds:
\begin{equation}
\begin{aligned}
&(h+h_E)\cdot q_A\cdot q_B\approx h_E\cdot q_A\cdot q_B\\
=&\left(q_B\cdot\mathbf{s}_{BE}\right)^T\cdot\mathbf{D}^T\cdot diag(\mathbf{w})\cdot\mathbf{D}\cdot\left(\mathbf{s}_{AE}\cdot q_A\right)
\end{aligned}
\end{equation} 
This envisages us to estimate the combined $\mathbf{s}_{AE}\cdot q_A$ ($\mathbf{s}_{BE}\cdot q_B$) from Eq. (\ref{cs3}). Given the sparse representations, i.e., $\|\mathbf{s}_{AE}\cdot q_A\|_0=\|\mathbf{s}_{AE}\|_0=\iota<5$ and $\|\mathbf{s}_{BE}\cdot q_B\|_0=\|\mathbf{s}_{BE}\|_0=\iota<5$, the compressed sensing problem can be built as:
\begin{equation}
\label{cs4}
\begin{aligned}
    &\min_{\mathbf{s}_{AE}\cdot q_A}\left\|\mathbf{r}_E^{(A)}-\mathbf{C}\mathbf{D}\cdot(\mathbf{s}_{AE}\cdot q_A)\right\|_2^2,~s.t.,~\|\mathbf{s}_{AE}\cdot q_A\|_0\leq \iota,\\
    &\min_{\mathbf{s}_{BE}\cdot q_B}\left\|\mathbf{r}_E^{(B)}-\mathbf{C}\mathbf{D}\cdot(\mathbf{s}_{BE}\cdot q_B)\right\|_2^2,~s.t.,~\|\mathbf{s}_{BE}\cdot q_B\|_0\leq \iota,
\end{aligned}
\end{equation}
where OMP is used to find the optimal (sub-optimal) solutions, denoted as $\widehat{\mathbf{s}_{AE}\cdot q_A}$ and $\widehat{\mathbf{s}_{BE}\cdot q_B}$. 

Then, Eve-RIS can reconstruct part of $\hat{\phi}_A$ (or $\hat{\phi}_B$) as:
\begin{equation}
\label{B18}
\hat{\phi}_E=\left(\widehat{\mathbf{s}_{BE}\cdot q_B}\right)^T\cdot\mathbf{D}^T\cdot diag(\mathbf{w})\cdot\mathbf{D}\cdot\left(\widehat{\mathbf{s}_{AE}\cdot q_A}\right)
\end{equation}
After the derivation of the feature $\hat{\phi}_E$, Eve-RIS can regenerate the legitimate secret key of Alice and Bob. This is achieved by computing $\hat{\varphi}_E=Re[\hat{\phi}_E]$ and replacing $z_E$ in Eq. (\ref{key}), given the shared information between $\hat{\phi}_E$ and $\hat{\phi}_A$, i.e., $h_E\cdot q_A\cdot q_B$.

{\color{blue}
\subsection{Eve-RIS Phase Optimization}
From the key match rate analysis of the proposed Eve-RIS designs in Eq. (\ref{theo1_1}), a better eavesdropping performance can be achieved by maximizing the variance of the Eve-RIS generated deceiving channel, i.e., $\sigma_E^2$. This is done by finding the optimized RIS phase vector $\mathbf{w}$. According to the expression of $\sigma_E^2$ in Eq. (\ref{eq4}), the optimization problem is formulated as:
\begin{equation}
\label{obj1}
\begin{aligned}
&\max_{\mathbf{w}}\mathbf{w}^H\mathbf{G}\mathbf{w},\\
&~s.t.,~\mathbf{w}^H\mathbf{E}_m\mathbf{w}=A_E,~m=1,\cdots,M
\end{aligned}
\end{equation}
where $\mathbf{E}_m\triangleq\mathbf{e}_m\cdot\mathbf{e}_m^H$, and $\mathbf{e}_m$ denotes a unit-norm vector whose $m$th element is $1$.

It is seen from Eq. (\ref{obj1}) that the optimization problem is with $M$ non-convex constraints, which makes it difficult to solve. To tackle this issue, we transform Eq. (\ref{obj1}) with an equivalent convex problem by relaxing its equality constraints, i.e., 
\begin{equation}
\label{obj2}
\begin{aligned}
&\max_{\mathbf{w}}\mathbf{w}^H\mathbf{G}\mathbf{w},\\
&~s.t.,~\mathbf{w}^H\mathbf{E}_m\mathbf{w}\leq A_E,~m=1,\cdots,M.
\end{aligned}
\end{equation}
The equivalency can be proved in the following. Suppose the optimal $\mathbf{w}$ is within the area of $\mathbf{w}^H\mathbf{E}_m\mathbf{w}<A_E$, say $\mathbf{w}_{opt}^H\mathbf{E}_m\mathbf{w}_{opt}<A_E$. Then, we assign $\mathbf{w}=\sqrt{A_E}\mathbf{w}_{opt}/\sqrt{\mathbf{w}_{opt}^H\mathbf{E}_m\mathbf{w}_{opt}}$, which has (i) a larger value of the objective function, i.e., $\mathbf{w}^H\mathbf{G}\mathbf{w}=A_E/(\mathbf{w}_{opt}^H\mathbf{E}_m\mathbf{w}_{opt})\mathbf{w}_{opt}^H\mathbf{G}\mathbf{w}_{opt}>\mathbf{w}_{opt}^H\mathbf{G}\mathbf{w}_{opt}$, and (ii) satisfaction of constraints, i.e., $\mathbf{w}^H\mathbf{E}_m\mathbf{w}=A_E$. This means the optimal $\mathbf{w}$ locates at the boundary, which therefore proves the equivalency between Eqs. (\ref{obj1})-(\ref{obj2}). 

From Eq. (\ref{obj2}), we transform the non-convex problem into the convex one, and the optimal solution can be obtained with the aid of
the CVX MATLAB toolbox.}

\section{Difference from Current Eavesdroppers}
In this section, we compare our designed Eve-RIS with other popular attackers, from the conceptual perspectives. Here, we categorize the attackers by whether the channel reciprocity between Alice and Bob is maintained or destroyed.

{\color{blue}
\subsection{Distinguish with Attackers Maintaining Channel Reciprocity}
The attackers that maintain the channel reciprocity, aim at partially estimating secret keys, by generating and inserting a reciprocal part to Alice's and Bob's channel probing results. In a mathematical view, the channel estimation results at Alice, Bob, and the attacker, denoted as $\hat{\psi}_A$, $\hat{\psi}_B$ and $\hat{\psi}_E$, are expressed as:
\begin{equation}
\begin{aligned}
\hat{\psi}_A=&h+\psi_E+\hat{n}_A,\\
\hat{\psi}_B=&h+\psi_E+\hat{n}_B,\\
\hat{\psi}_E=&\psi_E+\hat{n}_E,
\end{aligned}
\end{equation}
where $\psi_E$ is the inserted part from the attackers. Here, we compare our proposed Eve-RIS with other three popular types of attackers maintaining channel reciprocity, i.e.,
\begin{equation}
\label{att1}
\psi_E=
\begin{cases}
h_E & \text{Eve-RIS (proposed)}\\
\tilde{\mathbf{g}}_{BE}^T\cdot diag(\tilde{\mathbf{w}})\cdot\tilde{\mathbf{g}}_{AE} & \text{Untrust relay in \cite{thai2015physical,letafati2020new}}\\
1_{\|\tilde{\mathbf{g}}_{AE}\|_2\approx\|\tilde{\mathbf{g}}_{BE}\|_2}\cdot p & \text{Spoofing in \cite{eberz2012practical,pan2021man,8626511}}\\
\tilde{\mathbf{g}}_{BE}^T\cdot\mathbf{p}=\tilde{\mathbf{g}}_{AE}^T\cdot\mathbf{p} & \text{Spoofing in \cite{letafati2021deep}}
\end{cases}
\end{equation}
where $\tilde{\mathbf{g}}_{AE}$ and $\tilde{\mathbf{g}}_{BE}$ are the channels from Alice and Bob to the attacker, respectively (note that these channels are different from those of Alice and Bob to RIS, given the structural difference between RIS and other devices).

\subsubsection{Attackers with Physical Reciprocity}
Attackers that can physically maintain the channel reciprocity include our proposed Eve-RIS, and untrust relays in \cite{thai2015physical,letafati2020new}, whose combined channels (e.g., Alice to Eve-RIS/relay to Bob, and Bob to Eve-RIS/relay to Alice channels) naturally maintain the reciprocal property. Similar to the proposed Eve-RIS, the untrust relay can assign its transit vector $\tilde{\mathbf{w}}$ in Eq. (\ref{att1}) to insert a deceiving channel $\psi_E$, which enables to obtain the legitimate secret keys.

The main differences between our proposed Eve-RIS and untrust relays are in the following two aspects:

\begin{itemize}
\item Untrust relay attack is vulnerable to the existing relay transmission protocols \cite{thai2015physical,letafati2020new}. Such protocols request relays to send a pilot in the first place (e.g., phase 1 in \cite{thai2015physical}, and step 1 in \cite{letafati2020new}). Then, leveraging the estimation of channels from relays to legitimate users, a more secured PL-SKG can be designed to generate secret keys with no leakage, which therefore relieves the threat from the untrust relay and its performed MiM channel insertion attack.
By contrast, the defensive approaches designed for untrust relays cannot be implanted on our proposed Eve-RIS, as is less practical to assume a reflective surface to actively send pilots for protocol and authentication purposes. As such, the proposed Eve-RIS provides a better way to realize the MiM channel insertion attack, which therefore demonstrates an arising new eavesdropping threat that requires further countermeasure designs specific to Eve-RIS.

\item From the implementation view, both the untrust relay and our proposed Eve-RIS suffer from severe cascaded channel attenuation. For the untrust relay, the variance of its inserted channel in Eq. (\ref{att1}) is computed as $\|\tilde{\mathbf{w}}\|_2^2C_0^2d_{AE}^{-\alpha_L}d_{BE}^{-\alpha_L}$. This suggests that the only way to counter the cascaded attenuation is to increase the relaying (amplifying) gain, i.e., $\|\tilde{\mathbf{w}}\|_2^2$ (e.g., in Fig. \ref{figs2_3}, a $60$dB gain is required for a $25$m attack). In contrast with the untrust relay, such an amplifying gain is impractical even with active RIS elements. Alternatively, we leverage the advantages of RIS, i.e., can be equipped with more reflective elements, and its ability to optimize channel, and achieve a comparable eavesdropping performance with untrust relays at the expanse of low energy cost (see Fig. \ref{figs2_3}, Eve-RIS with $1600$ passive elements reaches equivalent performance with the untrust relay using $60$dB gain). 
\end{itemize}

\subsubsection{Attackers with Probabilistic Reciprocity}
Attackers that have a probability to maintain channel reciprocity include the works in \cite{eberz2012practical,pan2021man,8626511}. In these works, the attackers firstly estimate the RSS-based channels from Alice and Bob to them, i.e., $\|\tilde{\mathbf{g}}_{AE}\|_2^2$ and $\|\tilde{\mathbf{g}}_{BE}\|_2^2$, via Alice's and Bob's sending pilots. Then, the attacker waits for the chance of RSS-based channels reaching reciprocity, i.e., $\|\tilde{\mathbf{g}}_{AE}\|_2^2\approx\|\tilde{\mathbf{g}}_{BE}\|_2^2$, and send spoofing signal to both Alice and Bob (Eq. (4) in \cite{pan2021man}). As such, the RSS-based channel probing results at Alice and Bob contain the attacker-generated reciprocal parts, i.e., $p=\|\tilde{\mathbf{g}}_{AE}\|_2^2\approx\|\tilde{\mathbf{g}}_{BE}\|_2^2$, which can be used by the attacker to partially obtain the legitimate secret keys. 

The differences between our proposed Eve-RIS and these attackers are in the two aspects. 

\begin{itemize}
\item The attackers with probabilistic reciprocity are harshly restricted by an "attack opportunity", i.e., $Pr\{\|\tilde{\mathbf{g}}_{AE}\|_2^2-\|\tilde{\mathbf{g}}_{BE}\|_2^2<\epsilon\}$, only within which the attack can be pursued. This suggests the inability to eavesdropping the secret keys over a long duration period. By contrast, our proposed Eve-RIS can physically maintain the channel reciprocity, thereby enabling it to pursue the estimation of the legitimate secret keys continuously, without the limitation of "attack opportunity". 

\item The measuring of insert opportunity requires the exact channel estimation of $\|\tilde{\mathbf{g}}_{AE}\|_2$ and $\|\tilde{\mathbf{g}}_{BE}\|_2$. This suggests the vulnerability of these attackers to channel randomization, i.e., the two-way cross multiplication PL-SKG in Eq. (5) of \cite{8626511}, by sending random pilots to disable the channel estimations at the attacker. By comparison, one distinguish of our proposed Eve-RIS is its ability to defeat the two-way PL-SKG (in Section III. B), which thereby constitutes the difference from the defensive point of view. 
\end{itemize}

\subsubsection{Attackers with Created Reciprocity}
In Eq. (\ref{att1}), the spoofing scheme in \cite{letafati2021deep} is to create and send artificial reciprocal signals to Alice and Bob, i.e., $\tilde{\mathbf{g}}_{BE}^T\cdot\mathbf{p}$ ($\tilde{\mathbf{g}}_{AE}^T\cdot\mathbf{p}$), respectively in the Alice's and Bob's pilot sending time slots. Here, the precoding $\mathbf{p}$ is designed by the attacker to maintain the channel reciprocity, i.e., $\tilde{\mathbf{g}}_{BE}^T\cdot\mathbf{p}=\tilde{\mathbf{g}}_{AE}^T\cdot\mathbf{p}$.

Given the aforementioned process, the differences between our proposed Eve-RIS and this attack can be categorized as two aspects. 

\begin{itemize}
\item In the spoofing scheme \cite{letafati2021deep}, the precoding design, i.e., $\mathbf{p}$, requires exact estimations of channels from Alice and Bob to them, i.e., $\tilde{\mathbf{g}}_{BE}^T$ and $\tilde{\mathbf{g}}_{AE}$. This suggests its vulnerability if Alice and Bob use random pilots, i.e., the two-way based PL-SKG, which will ruin the attacker's channel probing process. This thereby provides a difference from the defensive perspective, as our designed Eve-RIS can defeat the two-way based PL-SKG (shown in Section III. B). 

\item Even if ordinary pilots are used by Alice and Bob, the attacker in \cite{letafati2021deep} is still difficult to obtain channel estimations of $\tilde{\mathbf{g}}_{AE}$ and $\tilde{\mathbf{g}}_{BE}$ in time to design its precoding $\mathbf{p}$. 
Consider the real-time scenario, where the channels from Alice and Bob to the spoofing Eve, i.e., $\tilde{\mathbf{g}}_{AE}$ and $\tilde{\mathbf{g}}_{BE}$, change independently for two consecutive channel estimation round (one channel estimation round contains an odd and an even time slot for Alice and Bob sending pilots in the TDD mode for channel estimations). In this view, when the spoofing Eve sends pilots in the same odd time slot as Alice, the spoofing Eve is hard to obtain the Bob-to-Eve channel of the current channel estimation round, i.e., $\tilde{\mathbf{g}}_{BE}$, since currently there is no pilot from Bob in the odd time slot (but will be in the following even time slot). As such, the precoding $\mathbf{p}$ to maintain channel reciprocity, i.e., $\psi_E=\tilde{\mathbf{g}}_{BE}^T\cdot\mathbf{p}=\tilde{\mathbf{g}}_{AE}^T\cdot\mathbf{p}$, cannot be generated in time for the attack in current channel estimation round. This, therefore, hinders the spoofing attack in \cite{letafati2021deep} to obtain the legitimate secret keys relying on the fast time-varying and independent CSI.  
\end{itemize}
}

\subsection{Distinguish with Attackers Destroying Channel Reciprocity}\label{with_spoofing}
The attackers that destroy the channel reciprocity include a wide range, e.g., pilot spoofing in \cite{zhou2012pilot,im2015secret}, and jamming \cite{9625442}. 
Here, we only compare with the pilot spoofing, as the jamming attackers are not designed to obtain legitimate secret keys. 
In the existence of a pilot spoofing Eve, the channel probing results at Alice and Bob are not reciprocal, i.e., \cite{zhou2012pilot,im2015secret}
\begin{equation}
\label{spoofing}
\begin{aligned}
    &\hat{\psi}_A=h+\tilde{g}_{AE}\frac{\sqrt{E_s}}{\|\mathbf{x}_B\|_2}+\hat{n}_A,\\
    &\hat{\psi}_B=h+\hat{n}_B,\\
    &\hat{\psi}_E^{(A)}=\tilde{g}_{AE}+\hat{n}_E^{(A)},\\
\end{aligned}
\end{equation}
where $\tilde{g}_{AE}\sim\mathcal{CN}(\tilde{g}_{AE}^\text{(LoS)},2\sigma_{AE}^2)$ is the single-antenna based channels from Alice and Bob to spoofing Eve. $E_S/\|\mathbf{x}_B\|_2^2$ is the spoofing gain of the legitimate pilots. Here, different from the spoofing in \cite{letafati2021deep} that aims to maintain channel reciprocity, the spoofing Eve in Eq. (\ref{spoofing}) aims to pretend as one of the legitimate users (e.g., Bob) and generates secret keys with another (e.g., Alice). This is achieved by increasing the spoofing gain $E_s/\|\mathbf{x}_B\|_2^2$, so $\tilde{g}_{AE}$ will dominate the channel probing result at Alice, making $\hat{\psi}_A$ and $\hat{\psi}_E^{(A)}$ alike.

The main difference between our designed Eve-RIS and the spoofing Eve in \cite{zhou2012pilot,im2015secret} is whether the legitimate channel is still reciprocal. In spoofing Eve scenarios, the channels between Alice and Bob are not reciprocal, due to the participation of the spoofing activity, i.e., $\hat{\psi}_A\neq\hat{\psi}_B$. In this view, Alice and Bob can compare their channel estimation results to determine whether a spoofing Eve exists. Compared to the spoofing Eve, the channels between Alice and Bob under Eve-RIS are still reciprocal, i.e., $h_A=h_B=h+h_E$. This, to some extent, helps conceal the Eve-RIS, as Alice and Bob cannot detect a considerable difference from their channel estimation results. Further discussion and results will be pursued via simulations in Section V. C.

\section{Simulation Results}
In this section, we evaluate our designed Eve-RIS schemes. The model configuration is provided in the following. In a 3D space, Alice and Bob are located at $(0,0,0)$, $(0,50,0)$, with unit m. Eve (either the Eve-RIS or the peering attackers) is located at $(0,10,5)$, unless other specifications. The direct channels from Alice and Bob to Eve-RIS are modeled in Eq. (\ref{eq02}) according to \cite{9300189}, where a square structure of RIS is considered, i.e., $M_x=M_y$. Here, the referenced path loss is set as $C_0=-30$dB at the reference distance (i.e., $1$m), and the LoS and NLoS path loss exponents are $\alpha_L=2$ and $\alpha_N=3$. The number of paths is $\iota=5$ \cite{6847111}, where the first path is the LoS path and the rest $4$ paths are NLoS paths with random half-space elevation and azimuth angles independently and randomly distributed over $\mathcal{U}[-\pi/2,\pi/2]$. For the PL-SKG using channel estimation results, the pilot sequences are set as publicly known with $\|\mathbf{x}_A\|_2^2=\|\mathbf{x}_B\|_2^2=0.1$W. For the two-way PL-SKG method, we assign $q_A,q_B\sim\mathcal{CN}(0,1\text{W})$. The variance of the receiving noise is assigned as $\sigma_n^2=-110$dBW. 

For the designed Eve-RIS, the number of channel sensors and RF chains for compressed sensing based channel probing and feature extraction is set as $C=20$. We examine different groups of amplifier gain $A_E$ and the number of Eve-RIS elements $M$, where $A_E$ ranges from $0$dB (passive RIS) to $30$dB and $M=M_x\times M_y$ are selected from $\{64,100,400,1600\}$ \cite{9300189}.

\subsection{Performance of Eve-RIS against CSI-based PL-SKG}
\begin{figure}[!t]
\centering
\includegraphics[width=3.5in]{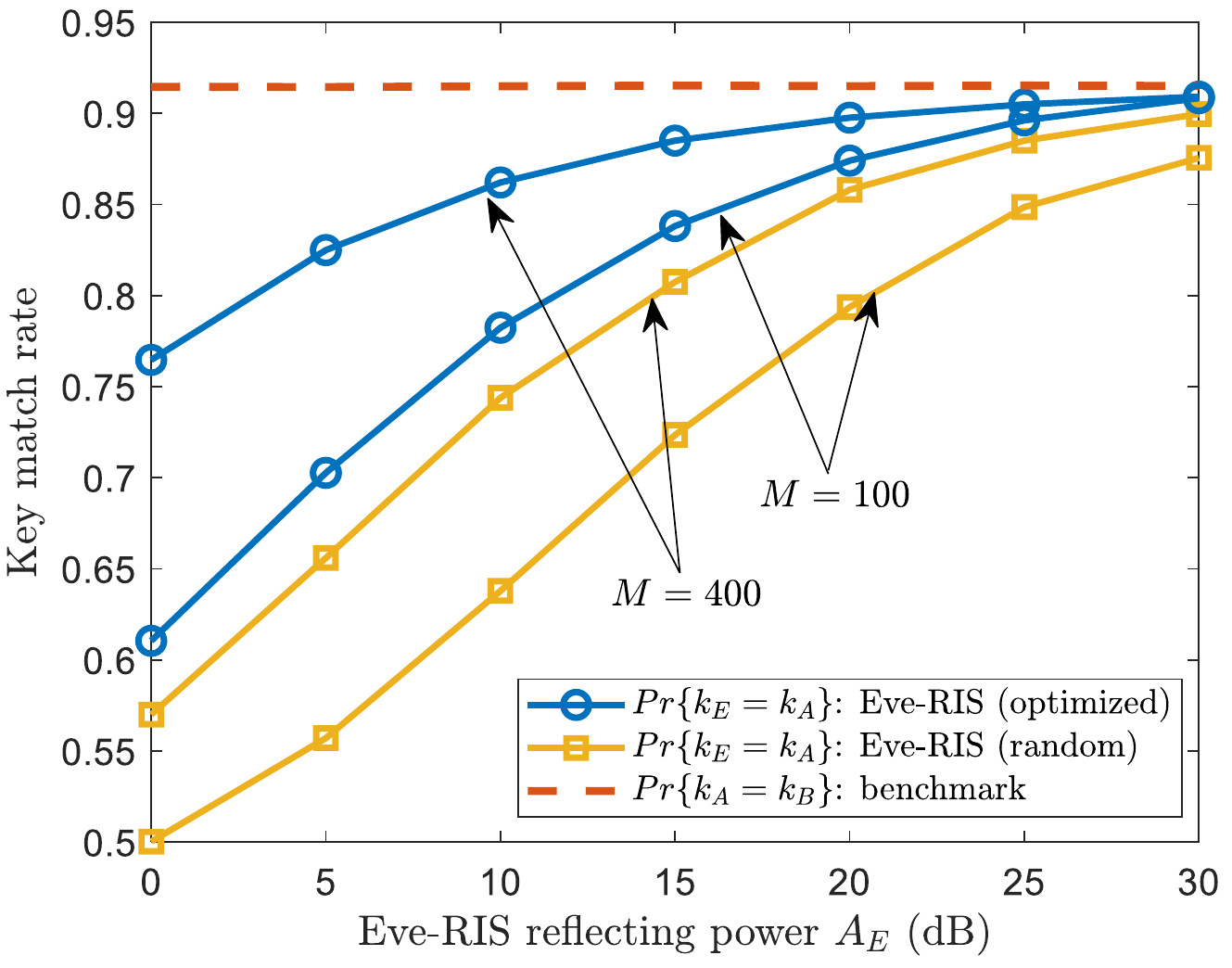}
\caption{Proposed Eve-RIS against CSI-based PL-SKG: Key match rate v.s. Eve-RIS amplifying gain.}
\label{figs1}
\end{figure}

\subsubsection{Key match rate analysis}
\begin{figure}[!t]
\centering
\includegraphics[width=3.5in]{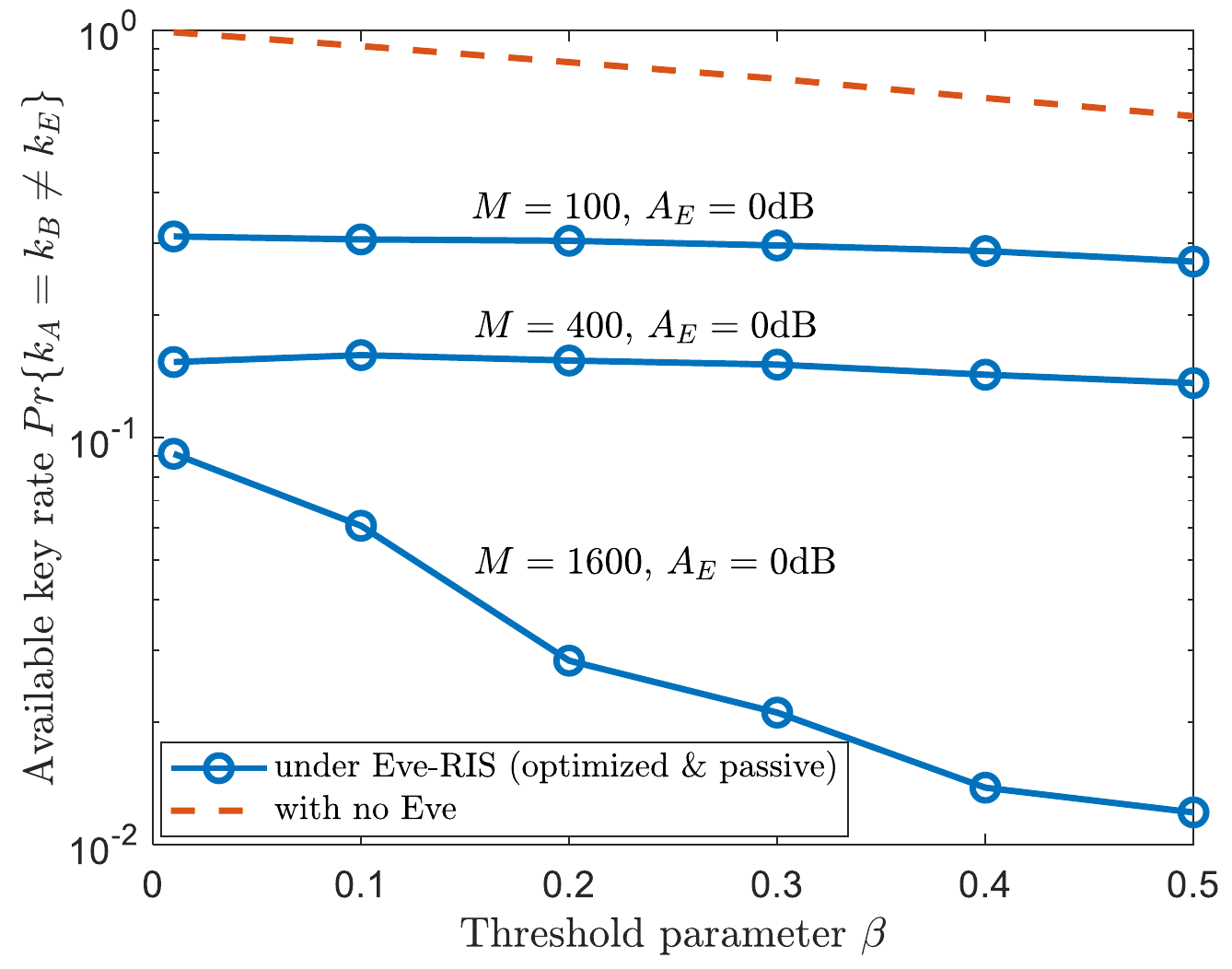}
\caption{Proposed Eve-RIS against CSI-based PL-SKG: Legitimate available key rate v.s. threshold parameter.}
\label{figs2}
\end{figure}

We first evaluate the key match rate between Alice and our proposed Eve-RIS, when attacking the CSI-based PL-SKG. For Fig. \ref{figs1}, the quantization threshold parameter is as $\beta=0.1$. Further results for different $\beta$ are shown by Figs. \ref{figs2}. In Fig. \ref{figs1}, the x-coordinate represents the amplifying gain of the Eve-RIS, i.e., $A_E$, while the y-coordinate gives the key match rate. 

It is first seen that with the increase of the Eve-RIS amplifying gain $A_E$, the key match rate between Alice and Eve, i.e., $Pr\{k_E=k_A\}$, grows. When $A_E>20$dB, $Pr\{k_E=k_A\}$ even approaches to key match rate of Alice and Bob, i.e., $Pr\{k_E=k_A\}\approx Pr\{k_A=k_B\}=0.92$. Second, it is observed that with the same Eve-RIS amplifying gain, a larger number of Eve-RIS elements, i.e., $M$, leads to a higher $Pr\{k_E=k_A\}$. For example, with the optimized Eve-RIS phase, when $A_E=10$dB, $Pr\{k_E=k_A\}$ increases from $0.75$ to $0.85$ as $M$ grows from $100$ to $400$. 
The reason behind these two observations is that both the amplifying gain and the number of Eve-RIS elements determine the variance of its generated deceiving channel, i.e., $\sigma_E^2\propto A_E\cdot M$ given by Eq. (\ref{eq4}), which, if increased, will increase $Pr\{k_E=k_A\}$ as deduced and analyzed by the theoretical key match rate in Eq. (\ref{theo1_1}). 

Then, it is observed that Eve-RIS using an optimized RIS phase has a larger key match rate with the legitimate user. For example, an increase of $Pr\{k_A=K_E\}$ from $0.65$ to $0.85$ at $A_E=5$dB is obtained by the optimal RIS phase. This is attributed to the RIS's ability to manipulate channels, which maximizes the variance of its generated deceiving channel and thereby achieves a better eavesdropping key match rate. 

\subsubsection{Legitimate key available rate analysis}
We next define and test the available key rate between Alice and Bob under our designed Eve-RIS, i.e., $Pr\{k_A=k_B\neq k_E\}$. In Fig. \ref{figs2}, \textcolor{blue}{the x-coordinate is the quantization threshold parameter $\beta$} while the y-coordinate is the available key rate, i.e., $Pr\{k_A=k_B\neq k_E\}$. It is first seen that with the increase of the quantization threshold $\beta$, all the available key rates between Alice and Bob with and without our designed Eve-RIS decrease. This is because a larger $\beta$ leads to a larger upper quantization threshold $\gamma_1$ and a smaller lower quantization threshold $\gamma_0$, \textcolor{blue}{which reduces the total number of keys.}

Second, we show in Fig. \ref{figs2} that the optimized Eve-RIS with passive reflecting elements can drastically decrease the legitimate available key rate between Alice and Bob, i.e., $Pr\{k_A=k_B\neq k_E\}$. 
For instance, given a fixed threshold parameter as $\beta=0.2$, $Pr\{k_A=k_B\neq k_E\}$ decreases from $0.80$ (no Eve) to $0.3$ (passive Eve-RIS with $M=100$ elements). Such a legitimate key rate can be further reduced to $0.03$ by the Eve-RIS using a larger number of elements (i.e., $M=6400$). 
This is because the variance of the Eve-RIS generated deceiving channel, i.e., $\sigma_E^2$, can be increased by enhancing not only the reflecting gain $A_E$ but also the number of RIS elements $M$, shown by Eq. (\ref{eq4}). This further demonstrates the eavesdropping potential of our designed Eve-RIS: even a passive RIS can achieve a threatening secret key leakage attack.

\subsection{Performance of Eve-RIS against two-way based PL-SKG} 

\begin{figure}[!t]
\centering
\includegraphics[width=3.5in]{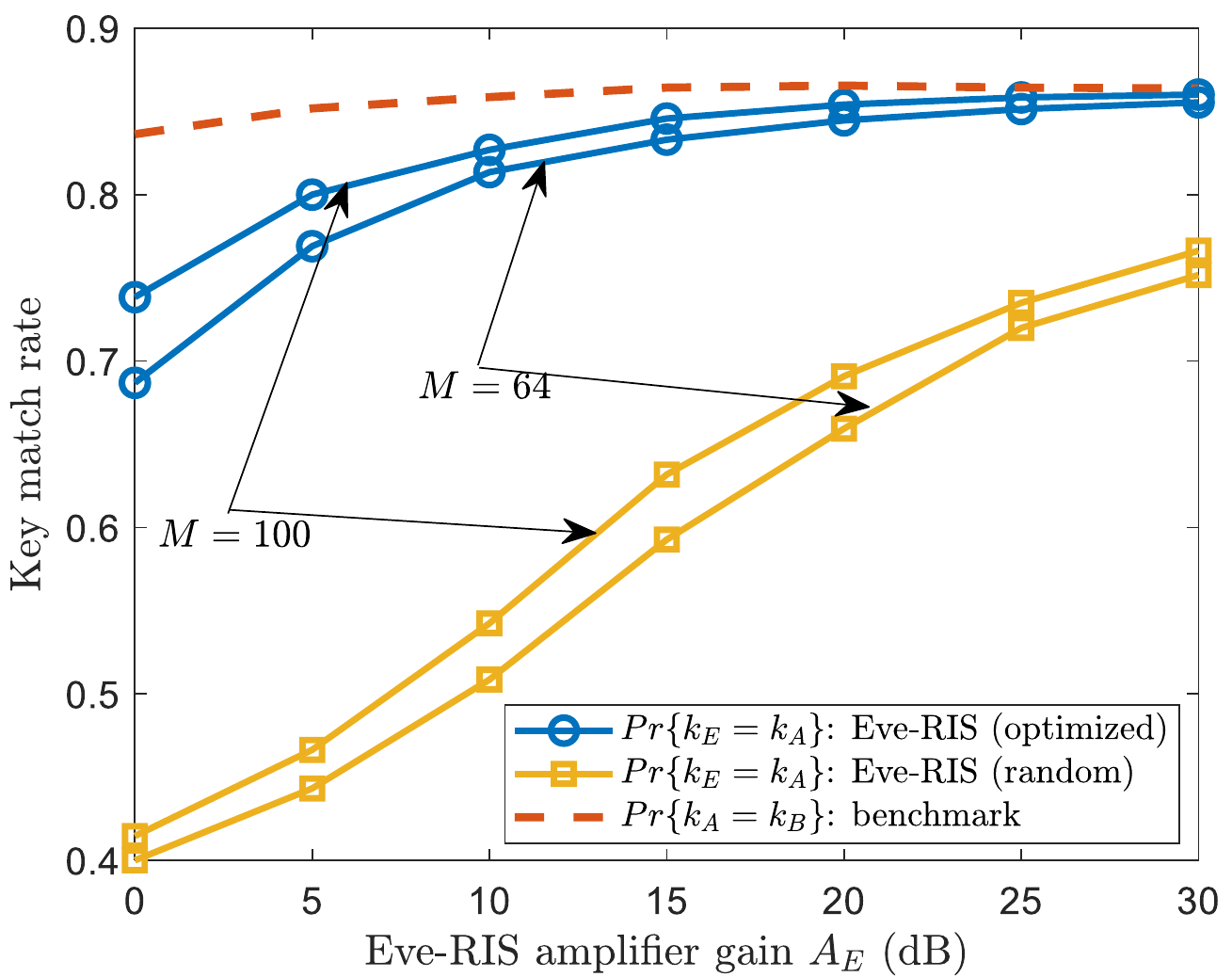}
\caption{Proposed Eve-RIS against two-way based PL-SKG: Key match rate v.s. Eve-RIS amplifying gain.}
\label{figs3}
\end{figure}

\begin{figure}[!t]
\centering
\includegraphics[width=3.5in]{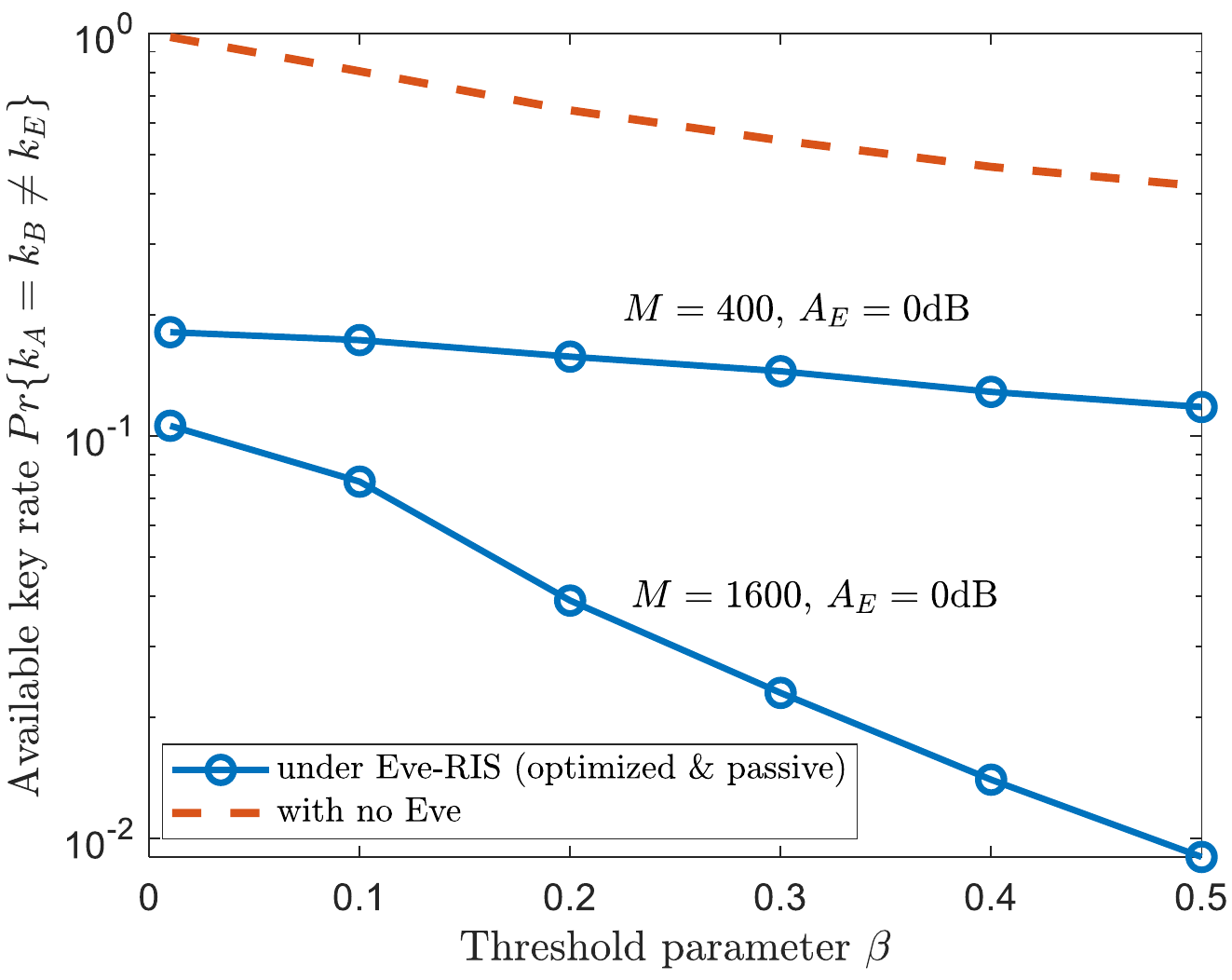}
\caption{Proposed Eve-RIS against two-way based PL-SKG: Available key rate v.s. quantization threshold parameter.}
\label{figs4}
\end{figure}

{\color{blue}We next evaluate the eavesdropping performance of our proposed Eve-RIS, when attacking two-way based PL-SKG. Similar results to Eve-RIS against CSI-based PL-SKG (i.e., Figs. \ref{figs1}-\ref{figs2}) can be seen in Figs. \ref{figs3}-\ref{figs4}. The proposed Eve-RIS with optimized phase has better eavesdropping ability (i.e., high key match rate $Pr\{k_A=K_E\}$ and small legitimate key rate $Pr\{k_A=k_B\neq k_E\}$), with the enhancement of either the number of RIS reflective elements $M$ or the reflecting gain $A_E$. Furthermore, it is observed from Fig. \ref{figs4} that a passive RIS (i.e., $M=6400,~A_E=0$dB) can also provide a promising attack effect against the two-way based PL-SKG, with a reduction of the legitimate key rate to $10^{-2}$.

The reason that our designed Eve-RIS can obtain the two-way PL-SKG based secret keys is different from the one against CSI-based PL-SKG, as the exact channel probing at Eve-RIS is unavailable due to the randomized channels by the two-way random pilots. Notably, neither the legitimate users nor the Eve-RIS relies on the original channels for key generation, but the features of randomized channels, i.e., $q_B(h+h_E)q_A$. For Eve-RIS, there is no information loss from the randomized Eve-RIS combined channel: the part of legitimate feature composed by this channel can be reconstructed with Eve-RIS's received signals, i.e., $q_Bh_Eq_A=(q_B\mathbf{g}_{BE})^T\cdot diag(\mathbf{w})\cdot(\mathbf{g}_{AE}q_A)$, which, therefore, guarantees the leakage attack of legitimate secret keys.}

\subsection{Comparison with Existing Attackers}\label{compare}
\textcolor{blue}{The comparison between our proposed Eve-RIS and other popular attackers mentioned in Section IV is pursued in this part. In the category of attackers that maintain channel reciprocity, we only select untrust relay for comparison, as other schemes (e.g., \cite{letafati2021deep,eberz2012practical,pan2021man,8626511}) strictly assume correlated time-varying channels in consecutive feature construction rounds, which, however, can be avoided by PL-SKGs using longer channel probing interval. For the attacker category that destroys the channel reciprocity, spoofing is selected for comparison, given the comparable aim to obtain the legitimate secret keys, other than jamming that is only interested in ruining the legitimate PL-SKG process. Also, given that the missing literature on how untrust relays and spoofing Eves obtain the two-way PL-SKG based secret keys, we only compare them by the attacking performance on CSI-based PL-SKG.}

\subsubsection{Detectability}
\begin{figure}[!t]
\centering
\includegraphics[width=3.5in]{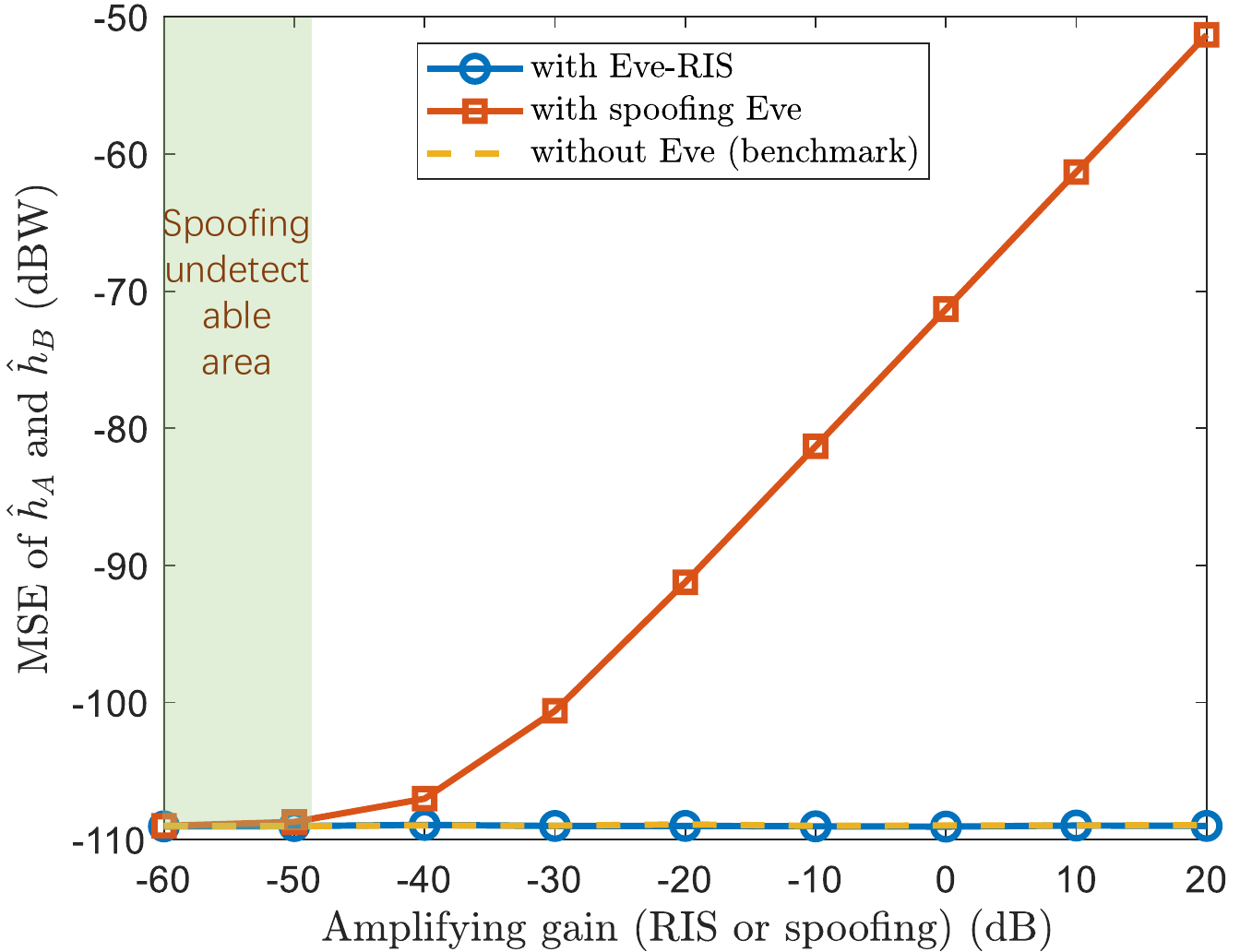}
\caption{Detectable comparison between the proposed Eve-RIS and the spoofing Eve. The proposed Eve-RIS can maintain the legitimate channel reciprocity and therefore avoid detection by Alice and Bob via their channel probing comparison.}
\label{figs5}
\end{figure}

In Fig. \ref{figs5}, we plot the mean square error (MSE) between Alice's and Bob's channel estimation results versus the amplifying gains of Eve (i.e., Eve-RIS's reflective gain, and the spoofing gain of legitimate pilots under the same quantitative level).
From Fig. \ref{figs5}, we see that for spoofing Eve, with the growth of the spoofing gain, the MSE between $\hat{h}_A$ and $\hat{h}_B$ increases drastically (e.g., from $-110$dBW to $-50$dBW), and is far from the benchmark, i.e., the MSE between $\hat{h}_A$ and $\hat{h}_B$ without an Eve. This suggests that the spoofing Eve can be easily detected by comparing the channel estimation results of Alice and Bob. \textcolor{blue}{Specially, a spoofing undetectable area is determined in Fig. \ref{figs5}, by pursuing a detection signal-to-noise ratio (SNR) on the MSE of $\hat{h}_A-\hat{h}_B$, i.e., $$\text{MSE}_\text{dBW}\leq\text{Benchmark floor}_\text{dBW}+\text{Detection SNR}_\text{dB},$$ which is $-110+1=-109$dBW in Fig. \ref{figs5}. As such, when the MSE is larger than $-109$dBW (corresponding to the spoofing amplifying gain larger than $-50$dB), the spoofing attacker can be detected via an energy detection method. }

{\color{blue}
Under our designed Eve-RIS, the difference between the channel probing results at Alice and Bob, (in terms of MSE between $\hat{h}_A$ and $\hat{h}_B$) remains the same as that of the benchmark (i.e., without an Eve). This is attributed to its ability to maintain channel reciprocity while pursuing a secret key leakage attack, which indicates the difficulty to be detected by checking the reciprocity of the legitimate channel probing results.

It is noteworthy that when Eve-RIS pursues a secret key attack, indeed a variance increase can be detected by one legitimate user. However, such a change can also happen with the environment changing, e.g., if two legitimate users are getting close to each other. In this view, detecting variance change is not sufficient to determine whether an Eve-RIS exists. Moreover, authentication for a RIS remains challenging. Unlike untrust relays that can actively send signals for online authentication purposes, it is difficult to pursue real-time authentication for Eve-RIS, given its inability to send authenticated messages via its reflective elements. Combing with these detectability challenges, and the resistance to current defensive approaches for untrust relays (discussed in Section IV. A), our proposed Eve-RIS presents a severe security threat, which therefore eagerly requires the design of novel targeted countermeasures to secure wireless data.}

\begin{figure}[!t]
\centering
\includegraphics[width=3.5in]{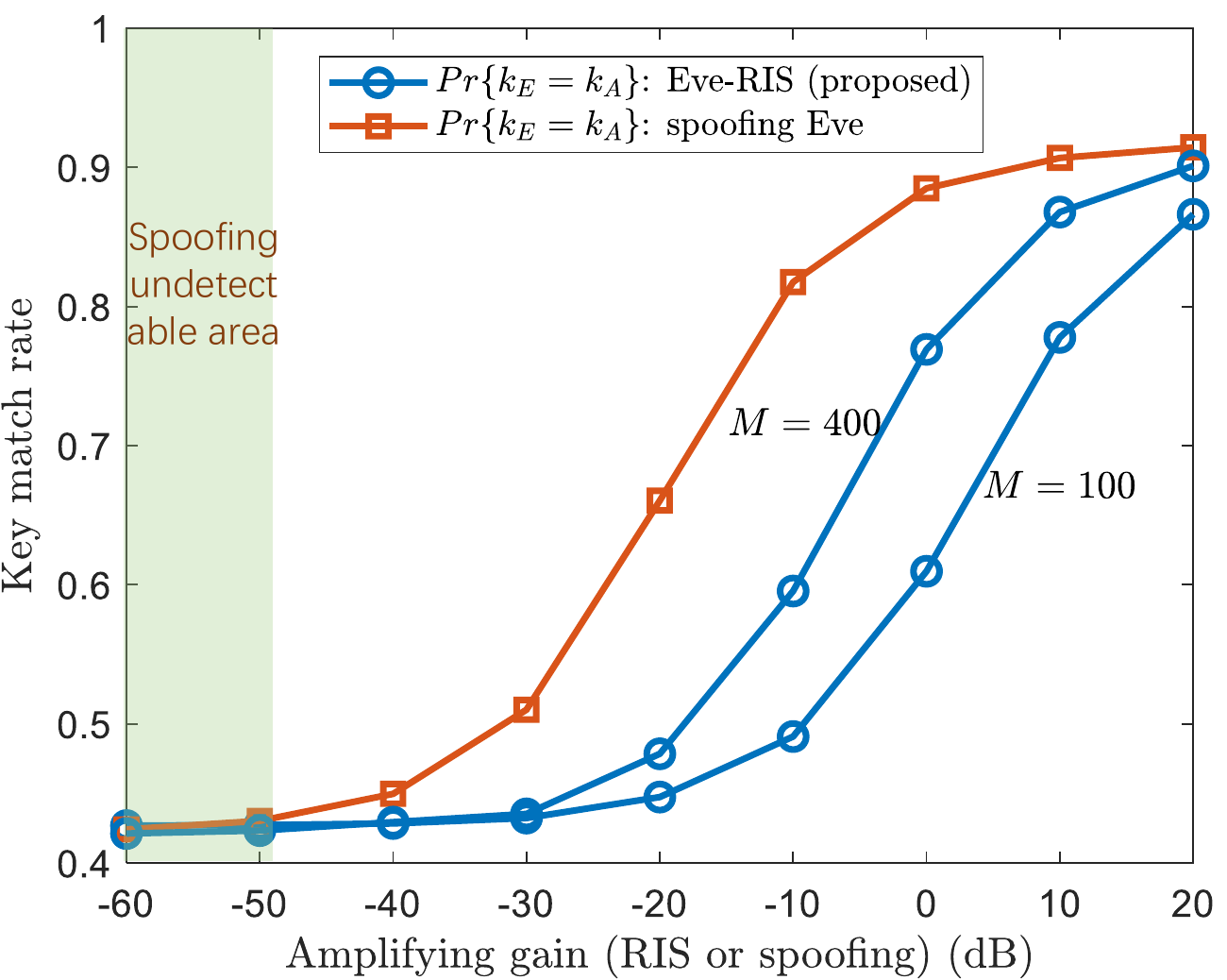}
\caption{Eavesdropping performance comparison between the proposed Eve-RIS, and the spoofing Eve.}
\label{figs6}
\end{figure}

\subsubsection{Eavesdropping ability}
We then compare the key match rates between our designed Eve-RIS and the spoofing Eve. In Fig. \ref{figs6}, the x-coordinate represents the amplifying gain (i.e., Eve-RIS's reflective gain, and the spoofing gain of legitimate pilots under the same quantitative level), and the y-coordinate is $Pr\{k_E=k_A\}$. 
It is first observed that compared with the spoofing Eve, our designed Eve-RIS requires either more reflective elements, i.e., $M$, or extra amplifying gains, i.e., $A_E$, to reach the same level of $Pr\{k_A=k_E\}$. This is because our proposed Eve-RIS has to compensate for the cascaded channel attenuation, i.e., $C_0^2d_{AE}^{-\alpha_L}d_{BE}^{-\alpha_L}$, while the spoofing Eve only needs to compensate for $C_0d_{AE}^{-\alpha_N}$. However, if we take into account the detectability for spoofing Eve, this means that (i) the available spoofing amplifying gain is $[-60\text{dB},-50\text{dB}]$, and (ii) the achievable key match rate of the spoofing Eve without being detected is less than $0.5$. By contrast, our designed Eve-RIS is difficult to be detected due to its ability to maintain the channel reciprocity between Alice and Bob, and therefore can have a promising key match rate without the concern about detection.

{\color{blue}
\begin{figure}[!t]
\centering
\includegraphics[width=3.5in]{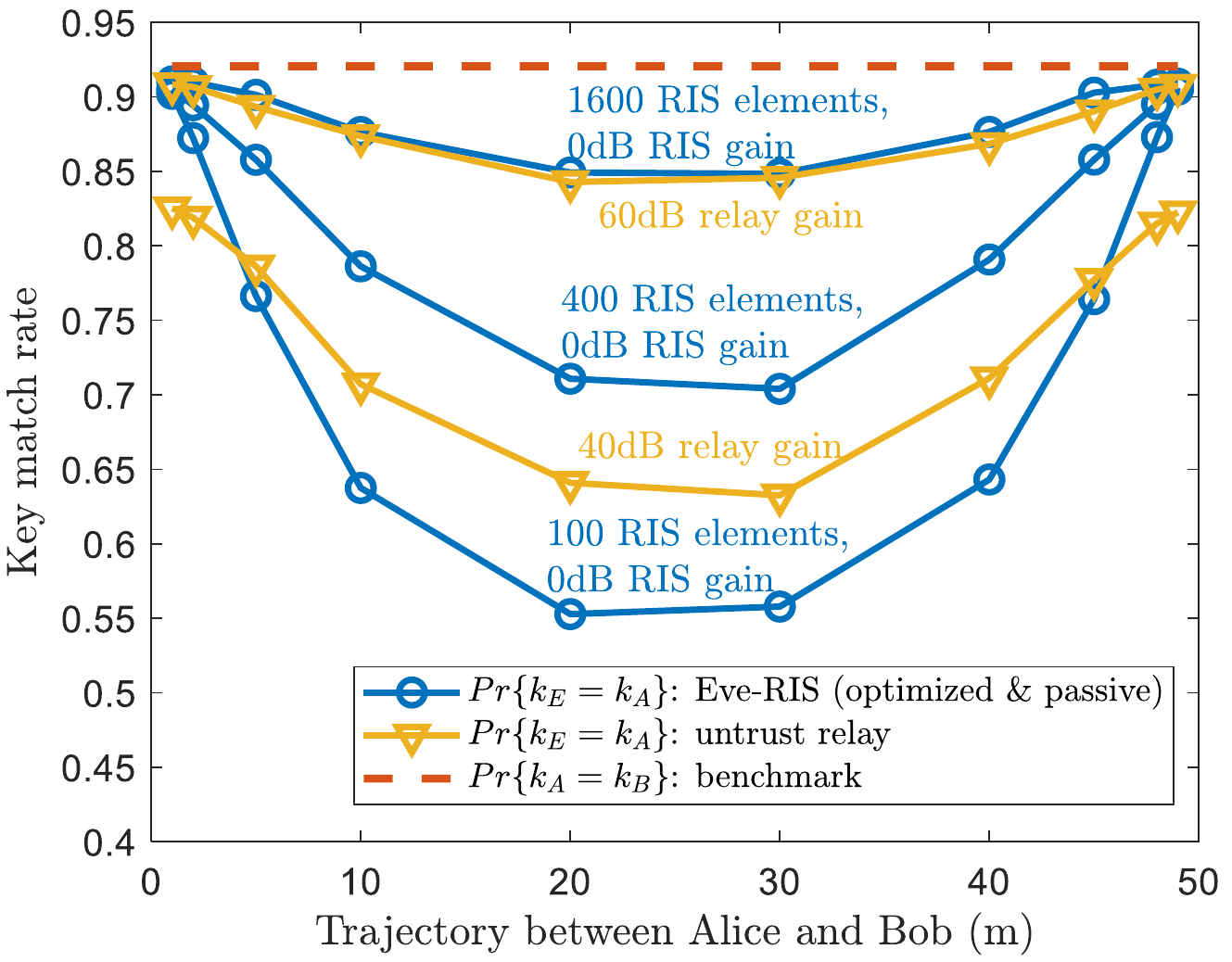}
\caption{Comparison between proposed Eve-RIS and untrust relays when attacking CSI-based PL-SKG. The x-coordinate represents the trajectory of the attacker from $(0,1,0)$ to $(0,49,0)$ (Alice and Bob are located at $(0,0,0)$ and $(0,50,0)$ with unit m), and y-coordinate is the key match rate. A comparable key match rate with the untrust relay using $60$dB gain can be seen by the Eve-RIS equipped with $1600$ passive elements. }
\label{figs2_3}
\end{figure}
\subsubsection{Performance with different Eve-RIS positions}
We further evaluate the key match rate between Eve (proposed Eve-RIS and the untrust relay) and the legitimate user, concerning different Eve positions. In Fig. \ref{figs2_3}, the x-coordinate represents the trajectory of Eve from $(0,1,0)$ to $(0,49,0)$ (Alice and Bob are located at $(0,0,0)$ and $(0,50,0)$ with unit m), and y-coordinate is the key match rate. As shown in Fig. \ref{figs2_3}, $3$ optimized Eve-RIS equipped with passive reflective elements are tested, where the number of RIS elements is selected from $M\in\{100,400,1600\}$. 

It is first seen that the key match rate $Pr\{k_A=K_E\}$ follows a symmetrical pattern over the trajectory: as Eve-RIS moves from Alice to Bob, $Pr\{k_A=K_E\}$ drops at first and then grows back after it passes the middle point, i.e., $(0,25,0)$. This is because when Eve-RIS is close to one legitimate user, the cascaded channel attenuation $C_0^2d_{AE}^{-\alpha_L}d_{BE}^{-\alpha_L}$ is smaller than when Eve-RIS is at the middle point, which therefore gives a larger variance of Eve-RIS's generated deceiving channel to obtain the legitimate secret keys. 

Then, it is observed that with the increase of the number of RIS elements $M$ (e.g., from $100$ to $1600$), $Pr\{k_A=K_E\}$ grows rapidly (e.g., from $0.55$ to $0.85$). This is because the variance of Eve-RIS generated and inserted channel can be increased by the enhancement of either the number of RIS elements $M$ or the RIS's reflecting gain $A_E$, as deduced in Eq. (\ref{eq4}). This thereby shows the eavesdropping potential of our proposed Eve-RIS, which, even equipped with passive reflective elements, can achieve a threatening secret key leakage attack.}

\textcolor{blue}{Third, a comparison with the untrust relay is shown in Fig. \ref{figs2_3}, where the proposed Eve-RIS equipped with $M=1600$ passive RIS elements reaches a comparable $Kr\{k_A=k_E\}$ to the untrust relay with $60$dB relaying gain. Recalling that the attacks from both the proposed Eve-RIS and the untrust relay suffer from the cascaded channel attenuation, which means the variances of their inserted deceiving channels should compensate for $C_0^2d_{AE}^{-\alpha_L}d_{BE}^{-\alpha_L}$. In contrast to the untrust relay that can only use amplifying gain for such compensation, our proposed Eve-RIS can leverage by (i) equipped with more RIS elements (as deduced in Eq. (\ref{eq4})), and (ii) the optimization of RIS phase to maximize the variance of its inserted channel. Also, given the resistance to the defensive approaches dealing with untrust relays (analyzed in Section IV. A 1)), our proposed Eve-RIS provides a new instance to implement the MiM channel insertion attack, which requires further specific countermeasure designs.}

{\color{blue}
\subsection{Discussion of Results and potential Countermeasures}
We here discuss the implementation and the potential countermeasures to our proposed Eve-RIS. To ensure a large inserted channel variance for secret key leakage attack, the implementation of the proposed Eve-RIS has two DoF, i.e., by either using a large number of RIS elements or increasing the amplifying gain. For example, when a structural constraint on RIS elements number is applied (e.g., $M=100$), the active RIS structure designed in \cite{ma2019controllable} may be adopted, which can provide a $20$dB amplifying gain (with corresponding $Pr\{k_A=k_E\}=0.86$ according to Fig. \ref{figs1}). On the other hand, such a level of $Pr\{k_A=k_E\}$ can also be achieved by a passive RIS with $M=1600$ elements \cite{9300189} (shown by Fig. \ref{figs2_3}). 

Then an open discussion on the potential countermeasures is provided. As studied in Section III, current PL-SKG based secret keys can be attacked and obtained by the proposed Eve-RIS. In this view, key-based PLS seems less attractive as a countermeasure or at least needs to be redesigned in the future works. Also, the defensive approaches for untrust relays that rely on relay transmission protocols are not suitable, given the inability of RIS to actively send signals via its reflective elements. Then, from key-less PLS, one potential method may be the beamforming of legitimate users to minimize the variance of Eve-RIS inserted channels. For this way, one should also consider the Eve-RIS anti-beamforming ability, given that the number of RIS elements is much larger than the number of antennas of Alice and Bob.}

\section{Conclusion}
In this paper, we demonstrated that the advance in RIS for securing the wireless communications is a double-edged sword. On the one hand, recent work has shown that RIS can improve the channel randomness and secrecy rate of PLS \cite{9520295,9201173, 9442833,9361290,9360860,staat2020intelligent}. On the other hand, our work here has shown that the presence of an adversarial Eve-controlled RIS (Eve-RIS) has the potential to reconstruct the PL-SKG based secret keys between Alice and Bob. We showed how the Eve-RIS can achieve this by generating and inserting a deceivingly random and reciprocal channel. As a result, current PL-SKGs with channel estimation and two-way cross-multiplication methods can be eavesdropped by our designed Eve-RIS scheme. 

Analysis and simulation results demonstrated the high key match rate obtained by our Eve-RIS with legitimate users, the low detectability as opposed to the spoofing Eve, and the resistance to most of the existing defensive approaches to untrust relays. As such, our proposed Eve-RIS provides a new eavesdropping threat on PL-SKG and should be seriously considered by further PL-SKG designs and security works in securing wireless communications.

\appendices

\section{Deduction of Eqs. (\ref{eq3})-(\ref{eq4})}\label{appendix1}
{\color{blue}
The detailed deduction of why Eve-RIS generated channel $h_E$ is approximated as complex Gaussian distributed is provided in the following. We first re-write the expression of $h_E$ from Eq. (\ref{eq1}), i.e., 
\begin{equation}
\label{eqA1}
h_E=\sum_{m=1}^Mw_m\cdot g_{AE,m}\cdot g_{BE,m},
\end{equation}
where $g_{AE,m}$ and $g_{BE,m}$ are the $m$th element of $\mathbf{g}_{AE}$ and $\mathbf{g}_{BE}$, respectively. In Eq. (\ref{eqA1}), $h_E$ is the summation of $M$ random variables with weak dependence, since $g_{AE,m}\cdot g_{BE,m}$ can only be independent with $g_{AE,n}\cdot g_{BE,n}$ when $n$th RIS element is half-wavelength away from $m$th RIS element \cite{9300189}. As such, given the central limit theorem under weak dependence (Theorem 27.5 in \cite{billingsley1995probability}), with a large number of RIS elements, e.g., $M>50$, $h_E$ can be approximated as complex Gaussian distribution, as shown in Eq. (\ref{eq3}).}

Next, we compute the mean and variance of $h_E$. Here, we re-write $h_E$ as the combinations of LoS and NLoS components, i.e., 
\begin{equation}
\label{eqA2}
\begin{aligned}
&h_E=\left[\mathbf{g}_{BE}^\text{(LoS)}+\mathbf{g}_{BE}^\text{(NLoS)}\right]^T\cdot diag(\mathbf{w})\cdot\left[\mathbf{g}_{AE}^\text{(LoS)}+\mathbf{g}_{AE}^\text{(NLoS)}\right]\\
=&\left(\mathbf{g}_{BE}^\text{(LoS)}\right)^T diag(\mathbf{w})\mathbf{g}_{AE}^\text{(LoS)}+\left(\mathbf{g}_{BE}^\text{(NLoS)}\right)^T diag(\mathbf{w})\mathbf{g}_{AE}^\text{(LoS)}\\
&+\left(\mathbf{g}_{BE}^\text{(LoS)}\right)^T diag(\mathbf{w})\mathbf{g}_{AE}^\text{(NLoS)}+\left(\mathbf{g}_{BE}^\text{(NLoS)}\right)^T diag(\mathbf{w})\mathbf{g}_{AE}^\text{(NLoS)}
\end{aligned}
\end{equation}
where $\mathbf{g}_{aE}^{\text{(LoS)}}$ and $\mathbf{g}_{aE}^{\text{(NLoS)}}$ ($a\in\{A,B\}$) represent the corresponding LoS and NLoS components, with relation $\mathbf{g}_{aE}=\mathbf{g}_{aE}^{\text{(LoS)}}+\mathbf{g}_{aE}^{\text{(NLoS)}}$. As such, the computation of mean and variance is determined by whether the RIS phase $\mathbf{w}$ is random or fixed (optimized), which will be discussed in the following respectively. 

\subsection{When RIS phase $\mathbf{w}$ is random}
In this case, all 4 terms in Eq. (\ref{eqA2}) are random variables, determined by either or combinations of the independent random phase, i.e., $\mathbf{w}$, and the NLoS channel components, i.e., $\mathbf{g}_{AE}^\text{(NLoS)}$ and $\mathbf{g}_{BE}^\text{(NLoS)}$. As such, the mean can be computed as:
\begin{equation}
\label{eqA3}
\begin{aligned}
&\mathbb{E}\left(h_E\right)\\
=&\left[\mathbf{g}_{BE}^\text{(LoS)}+\mathbb{E}\left(\mathbf{g}_{BE}^\text{(NLoS)}\right)\right]^Tdiag\left(\mathbb{E}(\mathbf{w})\right)\left[\mathbf{g}_{AE}^\text{(LoS)}+\mathbb{E}\left(\mathbf{g}_{AE}^\text{(NLoS)}\right)\right]\\
=&0,
\end{aligned}
\end{equation}
since (i) $\mathbb{E}(\mathbf{w})=\mathbf{0}$ when all $M$ RIS phases of $\mathbf{w}$ are independently and identically distributed over $\mathcal{U}[0,2\pi)$, and (ii) $\mathbb{E}[(\mathbf{g}_{AE}^{\text{(NLoS)}})^T\mathbf{g}_{BE}^{\text{(NLoS)}}]=\mathbb{E}(\mathbf{g}_{AE}^{\text{(NLoS)}})^T\mathbb{E}(\mathbf{g}_{BE}^{\text{(NLoS)}})$, given the in-dependency of $\mathbf{g}_{AE}^{\text{(NLoS)}}$ and $\mathbf{g}_{BE}^{\text{(NLoS)}}$ (as Alice and Bob are generally more than half-wavelength far from each other). 

The variance of $h_E$ is then computed by taking Eq. (\ref{eqA1}) and Eq. (\ref{eqA3}) into the its definition, i.e.,
\begin{equation}
\label{eqA4}
\begin{aligned}
&\mathbb{D}(h_E)=\mathbb{E}\left(h_E^*h_E\right)-\mathbb{E}\left(h_E^*\right)\mathbb{E}\left(h_E\right)=\mathbb{E}\left(h_E^*h_E\right)\\
=&\mathbb{E}\left(\left(\sum_{m=1}^Mw_m^*g_{AE,m}^*g_{BE,m}^*\right)\cdot\left(\sum_{m'=1}^Mw_{m'}g_{AE,m'}^*g_{BE,m'}^*\right)\right)\\
=&\sum_{m=1}^M\sum_{m'=1}^M\mathbb{E}\left(w_m^*w_{m'}\right)\mathbb{E}\left(g_{AE,m}^*g_{AE,m'}\right)\mathbb{E}\left(g_{BE,m}^*g_{BE,m'}\right)\\
\overset{(a)}{=}&\sum_{m=1}^M\mathbb{E}\left(|w_m|^2\right)\mathbb{E}\left(|g_{AE,m}|^2\right)\mathbb{E}\left(|g_{BE,m}|^2\right)\\
=&A_E\sum_{m=1}^M\left(2\Sigma_{AE,m,m}+\left|g_{AE,m}^\text{(LoS)}\right|^2\right)\left(2\Sigma_{BE,m,m}+\left|g_{BE,m}^\text{(LoS)}\right|^2\right)\\
\overset{(b)}{\approx}&A_E\cdot M\cdot C_0^2\cdot d_{AE}^{-\alpha_L}\cdot d_{BE}^{-\alpha_L}
\end{aligned}
\end{equation}
where $\Sigma_{AE,m,m}$ is the $(m,m)$th element of matrix $\bm{\Sigma}_{AE}$, and $\Sigma_{BE,m,m}$ is the $(m,m)$th element of matrix $\bm{\Sigma}_{BE}$.
In Eq. (\ref{eqA4}), $(a)$ is due to $\mathbb{E}(w_m^*w_{m'})=0$, given the independent random phase assignment for different RIS elements, i.e., $m\neq m'$. The approximation in $(b)$ is due to the fact that the energy of LoS component is greatly larger than that of NLoS component, e.g., $C_0d_{AE}^{\alpha_L}\gg C_0d_{AE}^{\alpha_N}$. From Eq. (\ref{eqA4}), $\sigma_E^2$ in Eq. (\ref{eq4}) can be computed by dividing 2. 

\subsection{When RIS phase $\mathbf{w}$ is fixed}
We then compute the mean and variance in the case RIS phase $\mathbf{w}$ is fixed, e.g., optimized RIS phase. From Eq. (\ref{eqA2}), if $\mathbf{w}$ is fixed, then the mean of $h_E$ is contributed by the cascaded LoS components, i.e., 
\begin{equation}
\begin{aligned}
\mathbb{E}\left(h_E\right)\!=&\!\left[\mathbf{g}_{BE}^\text{(LoS)}+\mathbb{E}\left(\mathbf{g}_{BE}^\text{(NLoS)}\right)\right]^T\!\!\!diag\left(\mathbf{w}\right)\left[\mathbf{g}_{AE}^\text{(LoS)}+\mathbb{E}\left(\mathbf{g}_{AE}^\text{(NLoS)}\right)\right]\\
=&\left(\mathbf{g}_{BE}^{\text{(LoS)}}\right)^T\cdot diag(\mathbf{w})\cdot\mathbf{g}_{AE}^{\text{(LoS)}},
\end{aligned}
\end{equation}
which thereby completes the computation of $\bm{\mu}_E$ in Eq. (\ref{eq3_4}).

We next compute the variance of $h_E$ with fixed RIS phase $\mathbf{w}$. It is seen from Eq. (\ref{eqA2}) that only the last 3 terms are random variables, determined by the NLoS channel components, i.e., $\mathbf{g}_{AE}^\text{(NLoS)}$ and $\mathbf{g}_{BE}^\text{(NLoS)}$. In this view, the variance can be computed as:
\begin{equation}
\label{33}
\begin{aligned}
&\!\!\!\!\mathbb{D}\left(h_E\right)\!=\!\mathbb{D}\bigg\{\!\!\Big[\!\!\left(\mathbf{g}_{AE}^\text{(NLoS)}\right)^T\!\!\!diag\left(\mathbf{g}_{BE}^\text{(LoS)}\right)\!+\!\left(\mathbf{g}_{BE}^\text{(NLoS)}\right)^T\!\!\!diag\left(\mathbf{g}_{AE}^\text{(LoS)}\right)\\
&~~~~~~~~~~~+\left(\mathbf{g}_{AE}^\text{(NLoS)}\odot\mathbf{g}_{BE}^\text{(NLoS)}\right)^T\Big]\mathbf{w}\bigg\}\\
=&\!\mathbf{w}^H\!\mathbb{E}\Bigg\{\!\bigg[\!diag\left(\mathbf{g}_{BE}^\text{(LoS)}\right)^*\!\!\left(\mathbf{g}_{AE}^\text{(NLoS)}\right)^*\!\!\!\!+\!diag\left(\mathbf{g}_{AE}^\text{(LoS)}\right)^*\left(\mathbf{g}_{BE}^\text{(NLoS)}\right)^*\!\!\!\\
&\!+\!\left(\mathbf{g}_{BE}^\text{(NLoS)}\odot\mathbf{g}_{AE}^\text{(NLoS)}\right)^*\!\!\bigg]\!\cdot\!
\bigg[\!\!\left(\mathbf{g}_{AE}^\text{(NLoS)}\right)^T\!\!\!diag\left(\mathbf{g}_{BE}^\text{(LoS)}\right)\\
&\!+\!\left(\mathbf{g}_{BE}^\text{(NLoS)}\right)^T\!\!\!diag\left(\mathbf{g}_{AE}^\text{(LoS)}\right)+\left(\mathbf{g}_{AE}^\text{(NLoS)}\odot\mathbf{g}_{BE}^\text{(NLoS)}\right)^T\bigg]\Bigg\}\mathbf{w}\\
=&\mathbf{w}^H\bigg\{diag\left(\mathbf{g}_{BE}^\text{(LoS)}\right)^*\!\!\!\mathbb{E}\left[\left(\mathbf{g}_{AE}^\text{(NLoS)}\right)^*\left(\mathbf{g}_{AE}^\text{(NLoS)}\right)^T\right]diag\left(\mathbf{g}_{BE}^\text{(LoS)}\right)\\
&+diag\left(\mathbf{g}_{AE}^\text{(LoS)}\right)^*\!\!\!\mathbb{E}\left[\left(\mathbf{g}_{BE}^\text{(NLoS)}\right)^*\left(\mathbf{g}_{BE}^\text{(NLoS)}\right)^T\right]diag\left(\mathbf{g}_{AE}^\text{(LoS)}\right)\\
&+\mathbb{E}\left[\left(\mathbf{g}_{BE}^\text{(NLoS)}\odot\mathbf{g}_{AE}^\text{(NLoS)}\right)^*\cdot\left(\mathbf{g}_{AE}^\text{(NLoS)}\odot\mathbf{g}_{BE}^\text{(NLoS)}\right)^T\right]\bigg\}\mathbf{w}\\
=&\mathbf{w}^H\bigg\{2diag\left(\mathbf{g}_{BE}^\text{(LoS)}\right)^*\!\!\!\bm{\Sigma}_{AE}diag\left(\mathbf{g}_{BE}^\text{(LoS)}\right)\\
&+2diag\left(\mathbf{g}_{AE}^\text{(LoS)}\right)^*\!\!\!\bm{\Sigma}_{BE}diag\left(\mathbf{g}_{AE}^\text{(LoS)}\right)+4\bm{\Sigma}_{AE}\odot\bm{\Sigma}_{BE}\bigg\}\mathbf{w}, 
\end{aligned}
\end{equation}
which, divided by $2$, gives the result in Eq. (\ref{eq4}) with fixed $\mathbf{w}$.

{\color{blue}
\section{Deduction of Theoretical Key Match Rate in Eq. (\ref{theo1_1})}\label{append2}
The theoretical key match rate in Eq. (\ref{theo1_1}) is deduced in the following. 
\begin{align}
    &Pr\{k_A=k_E\}\nonumber\\
    =&Pr\left(z_A>\gamma_1^\text{(A)},z_E>\gamma_1^\text{(E)}\right)+Pr\left(z_A<\gamma_0^\text{(A)},z_E<\gamma_0^\text{(E)}\right)\nonumber\\
    =&\iint_{\substack{\left\{\upsilon>\gamma_1^\text{(A)},\zeta>\gamma_1^\text{(E)}\right\}\\\bigcup\left\{\upsilon<\gamma_0^\text{(A)},\zeta<\gamma_0^\text{(E)}\right\}}}p_{z_A|z_E}(\upsilon|\zeta)\cdot p_{z_E}(\zeta)d\upsilon d\zeta\nonumber\\
    =&\iint_{\substack{\left\{\zeta+\xi>\gamma_1^\text{(A)},\zeta>\gamma_1^\text{(E)}\right\}\\\bigcup\left\{\zeta+\xi<\gamma_0^\text{(A)},\zeta<\gamma_0^\text{(E)}\right\}}}p_z(\xi)\cdot p_{z_E}(\zeta)d\xi d\zeta\nonumber\\
    =&\iint_{\substack{\left\{\zeta+\xi>\gamma_1^\text{(A)},\zeta>\gamma_1^\text{(E)}\right\}\\\bigcup\left\{\zeta+\xi<\gamma_0^\text{(A)},\zeta<\gamma_0^\text{(E)}\right\}}}\mathcal{N}(\xi, 0, \sigma_h^2)\cdot\mathcal{N}(\zeta, 0, \sigma_E^2)d\xi d\zeta\nonumber\\
    =&\frac{1}{\sqrt{2\pi}}\!\!\int_{\frac{\gamma_1^\text{(E)}}{\sigma_E}}^{+\infty}\!\!\Phi\left(\frac{-\gamma_1^\text{(A)}+\sigma_E\zeta}{\sigma_h}\right)\!\!\exp\left(-\frac{\zeta^2}{2}\right)d\zeta\nonumber\\
    &+\frac{1}{\sqrt{2\pi}}\!\!\int_{-\infty}^{\frac{\gamma_0^\text{(E)}}{\sigma_E}}\!\!\Phi\left(\frac{\gamma_0^\text{(A)}-\sigma_E\zeta}{\sigma_h}\right)\!\!\exp\left(-\frac{\zeta^2}{2}\right)d\zeta\nonumber\\
    =&\sqrt{\frac{2}{\pi}}\int_{\beta}^{+\infty}\Phi\left(-\beta\sqrt{\frac{\sigma_E^2}{\sigma_h^2}+1}+\frac{\sigma_E}{\sigma_h}\zeta\right)\exp\left(-\frac{\zeta^2}{2}\right)d\zeta
\end{align}
where $p_{z_A|z_E}(\upsilon|\zeta)$ is the PDF of $z_A$ conditioned on $z_E$, and $p_{z_E}(\zeta)$ is the PDF of $z_E$. $\mathcal{N}(\xi,0,\sigma^2)$ is the real Gaussian PDF of $\xi$ with $0$ expectation and $\sigma^2$ as variance.

Then, by assigning $x\triangleq\sigma_E/\sigma_h$, we compute the first-order derivative, i.e., $\partial{\Pr\{k_A=k_E\}}/\partial(x)$ to show its monotonically increasing property, i.e., 
\begin{equation}
\begin{aligned}
&\frac{\partial Pr\{k_A=k_E\}}{\partial x}\\
=&\frac{1}{\pi}\int_{\beta}^{+\infty}\!\!\!\!\!\!\!\exp\left(\!\!-\frac{\left(x\zeta\!-\!\beta\sqrt{x^2\!+\!1}\right)^2\!+\!\zeta^2}{2}\!\!\right)\!\!\!\left(\!\!\zeta\!-\!\frac{\beta}{\sqrt{1\!+\!\frac{1}{x^2}}}\!\!\right)d\zeta\overset{(a)}{>}0
\end{aligned}
\end{equation}
where $(a)$ is because (i) $\exp(-((x\zeta-\beta\sqrt{x^2+1})^2+\zeta^2)/2)>0$, and (ii) $\zeta-\beta/\sqrt{1+1/x^2}$ takes its minimum value $0$ when $x\rightarrow+\infty$ and $\zeta=\beta$.

Next, we prove $Pr\{k_A=k_E\}$ at first increases quickly and then gradually, with respect to $\sigma_E^2/\sigma_h^2$. This is done by evaluating its second-order derivative. As we denote $f_1(x,\zeta)\triangleq\exp(-((x\zeta-\beta\sqrt{x^2+1})^2+\zeta^2)/2)$ and $f_2(x,\zeta)\triangleq(\zeta-\beta/\sqrt{1+1/x^2})$, the second-order derivative can be computed as:
\begin{equation}
\begin{aligned}
&\frac{\partial^2 Pr\{k_A=k_E\}}{\partial x^2}=\frac{1}{\pi}\int_{\beta}^{+\infty}\frac{\partial \left[f_1(x,\zeta)\cdot f_2(x,\zeta)\right]}{\partial x}d\zeta\\
=&\frac{1}{\pi}\int_{\beta}^{+\infty}\!\!\!\!\!\!-f_1(x,\zeta)f_2^2(x,\zeta)\left(x\zeta-\beta\sqrt{x^2+1}\right)\\
&-f_1(x,\zeta)\frac{\beta}{(1+x^2)^{\frac{3}{2}}}d\zeta<0.
\end{aligned}
\end{equation}
}

\bibliographystyle{IEEEtran}
\bibliography{main.bib}

\begin{thebibliography}{10}
\providecommand{\url}[1]{#1}
\csname url@samestyle\endcsname
\providecommand{\newblock}{\relax}
\providecommand{\bibinfo}[2]{#2}
\providecommand{\BIBentrySTDinterwordspacing}{\spaceskip=0pt\relax}
\providecommand{\BIBentryALTinterwordstretchfactor}{4}
\providecommand{\BIBentryALTinterwordspacing}{\spaceskip=\fontdimen2\font plus
\BIBentryALTinterwordstretchfactor\fontdimen3\font minus
  \fontdimen4\font\relax}
\providecommand{\BIBforeignlanguage}[2]{{%
\expandafter\ifx\csname l@#1\endcsname\relax
\typeout{** WARNING: IEEEtran.bst: No hyphenation pattern has been}%
\typeout{** loaded for the language `#1'. Using the pattern for}%
\typeout{** the default language instead.}%
\else
\language=\csname l@#1\endcsname
\fi
#2}}
\providecommand{\BIBdecl}{\relax}
\BIBdecl

\bibitem{8883127}
H.-M. Wang, X.~Zhang, and J.-C. Jiang, ``{UAV-Involved Wireless Physical-Layer
  Secure Communications: Overview and Research Directions},'' \emph{IEEE
  Wireless Communications}, vol.~26, no.~5, pp. 32--39, 2019.

\bibitem{9520776}
W.~Wang, X.~Liu, J.~Tang, N.~Zhao, Y.~Chen, Z.~Ding, and X.~Wang,
  ``{Beamforming and Jamming Optimization for IRS-Aided Secure NOMA
  Networks},'' \emph{IEEE Transactions on Wireless Communications}, vol.~21,
  no.~3, pp. 1557--1569, 2022.

\bibitem{9656117}
X.~Pang, N.~Zhao, J.~Tang, C.~Wu, D.~Niyato, and K.-K. Wong, ``{IRS-Assisted
  Secure UAV Transmission via Joint Trajectory and Beamforming Design},''
  \emph{IEEE Transactions on Communications}, vol.~70, no.~2, pp. 1140--1152,
  2022.

\bibitem{8456560}
Y.~Zhou, P.~L. Yeoh, H.~Chen, Y.~Li, R.~Schober, L.~Zhuo, and B.~Vucetic,
  ``{Improving Physical Layer Security via a UAV Friendly Jammer for Unknown
  Eavesdropper Location},'' \emph{IEEE Transactions on Vehicular Technology},
  vol.~67, no.~11, pp. 11\,280--11\,284, 2018.

\bibitem{6739367}
A.~Mukherjee, S.~A.~A. Fakoorian, J.~Huang, and A.~L. Swindlehurst,
  ``{Principles of Physical Layer Security in Multiuser Wireless Networks: A
  Survey},'' \emph{IEEE Communications Surveys Tutorials}, vol.~16, no.~3, pp.
  1550--1573, 2014.

\bibitem{7393435}
J.~Zhang, T.~Q. Duong, A.~Marshall, and R.~Woods, ``{Key Generation From
  Wireless Channels: A Review},'' \emph{IEEE Access}, vol.~4, pp. 614--626,
  2016.

\bibitem{4036441}
C.~Ye, A.~Reznik, and Y.~Shah, ``{Extracting Secrecy from Jointly Gaussian
  Random Variables},'' in \emph{2006 IEEE International Symposium on
  Information Theory}, 2006, pp. 2593--2597.

\bibitem{5422766}
C.~Ye, S.~Mathur, A.~Reznik, Y.~Shah, W.~Trappe, and N.~B. Mandayam,
  ``{Information-Theoretically Secret Key Generation for Fading Wireless
  Channels},'' \emph{IEEE Transactions on Information Forensics and Security},
  vol.~5, no.~2, pp. 240--254, 2010.

\bibitem{5371757}
X.~Wu, Y.~Song, C.~Zhao, and X.~You, ``{Secrecy extraction from correlated
  fading channels: An upper bound},'' in \emph{2009 International Conference on
  Wireless Communications \& Signal Processing}, 2009, pp. 1--3.

\bibitem{7933224}
Y.~Peng, P.~Wang, W.~Xiang, and Y.~Li, ``{Secret Key Generation Based on
  Estimated Channel State Information for TDD-OFDM Systems Over Fading
  Channels},'' \emph{IEEE Transactions on Wireless Communications}, vol.~16,
  no.~8, pp. 5176--5186, 2017.

\bibitem{8293762}
K.~Moara-Nkwe, Q.~Shi, G.~M. Lee, and M.~H. Eiza, ``{A Novel Physical Layer
  Secure Key Generation and Refreshment Scheme for Wireless Sensor Networks},''
  \emph{IEEE Access}, vol.~6, pp. 11\,374--11\,387, 2018.

\bibitem{6171198}
S.~N. Premnath, S.~Jana, J.~Croft, P.~L. Gowda, M.~Clark, S.~K. Kasera,
  N.~Patwari, and S.~V. Krishnamurthy, ``{Secret Key Extraction from Wireless
  Signal Strength in Real Environments},'' \emph{IEEE Transactions on Mobile
  Computing}, vol.~12, no.~5, pp. 917--930, 2013.

\bibitem{mathur2008radio}
S.~Mathur, W.~Trappe, N.~Mandayam, C.~Ye, and A.~Reznik, ``{Radio-Telepathy:
  Extracting a Secret Key from an Unauthenticated Wireless Channel},'' in
  \emph{Proceedings of the 14th ACM international conference on Mobile
  computing and networking}, 2008, pp. 128--139.

\bibitem{10.1007/3-540-48285-7_35}
G.~Brassard and L.~Salvail, ``{Secret-Key Reconciliation by Public
  Discussion},'' in \emph{Advances in Cryptology --- EUROCRYPT '93},
  T.~Helleseth, Ed.\hskip 1em plus 0.5em minus 0.4em\relax Berlin, Heidelberg:
  Springer Berlin Heidelberg, 1994, pp. 410--423.

\bibitem{impagliazzo1989pseudo}
R.~Impagliazzo, L.~A. Levin, and M.~Luby, ``{Pseudo-Random Generation from
  One-Way Functions},'' in \emph{Proceedings of the twenty-first annual ACM
  symposium on Theory of computing}, 1989, pp. 12--24.

\bibitem{9000831}
N.~Aldaghri and H.~Mahdavifar, ``{Physical Layer Secret Key Generation in
  Static Environments},'' \emph{IEEE Transactions on Information Forensics and
  Security}, vol.~15, pp. 2692--2705, 2020.

\bibitem{lu2022secret}
X.~Lu, J.~Lei, and W.~Li, ``{Secret Key Generation based on Signal Power
  Allocation Optimisation},'' \emph{IET Communications}, vol.~16, no.~14, pp.
  1724--1730, 2022.

\bibitem{8254029}
G.~Li, A.~Hu, J.~Zhang, and B.~Xiao, ``{Security Analysis of a Novel Artificial
  Randomness Approach for Fast Key Generation},'' in \emph{GLOBECOM 2017 - 2017
  IEEE Global Communications Conference}, 2017, pp. 1--6.

\bibitem{lou2017secret}
Y.~Lou, L.~Jin, Z.~Zhong, K.~Huang, and S.~Zhang, ``{Secret Key Generation
  Scheme based on MIMO Received Signal Spaces},'' \emph{Scientia Sinica
  Informationis}, vol.~47, no.~3, pp. 362--373, 2017.

\bibitem{7582525}
H.~Taha and E.~Alsusa, ``{Secret Key Exchange Using Private Random Precoding in
  MIMO FDD and TDD Systems},'' \emph{IEEE Transactions on Vehicular
  Technology}, vol.~66, no.~6, pp. 4823--4833, 2017.

\bibitem{7593219}
A.~Khisti, ``{Secret-Key Agreement over Non-Coherent Block-Fading Channels with
  Public Discussion},'' \emph{IEEE Transactions on Information Theory},
  vol.~62, no.~12, pp. 7164--7178, 2016.

\bibitem{8057119}
S.~Sharifian, F.~Lin, and R.~Safavi-Naini, ``{Secret key agreement using a
  virtual wiretap channel},'' in \emph{IEEE INFOCOM 2017 - IEEE Conference on
  Computer Communications}, 2017, pp. 1--9.

\bibitem{8424614}
S.~Zhang, L.~Jin, Y.~Lou, and Z.~Zhong, ``{Secret Key Generation based on
  Two-Way Randomness for TDD-SISO System},'' \emph{China Communications},
  vol.~15, no.~7, pp. 202--216, 2018.

\bibitem{7794581}
G.~Wunder, R.~Fritschek, and K.~Reaz, ``{RECiP: Wireless Channel Reciprocity
  Restoration Method for Varying Transmission Power},'' in \emph{2016 IEEE 27th
  Annual International Symposium on Personal, Indoor, and Mobile Radio
  Communications (PIMRC)}, 2016, pp. 1--5.

\bibitem{9326394}
Q.~Wu, S.~Zhang, B.~Zheng, C.~You, and R.~Zhang, ``{Intelligent Reflecting
  Surface-Aided Wireless Communications: A Tutorial},'' \emph{IEEE Transactions
  on Communications}, vol.~69, no.~5, pp. 3313--3351, 2021.

\bibitem{8910627}
Q.~Wu and R.~Zhang, ``Towards smart and reconfigurable environment: Intelligent
  reflecting surface aided wireless network,'' \emph{IEEE Communications
  Magazine}, vol.~58, no.~1, pp. 106--112, 2020.

\bibitem{8811733}
------, ``{Intelligent Reflecting Surface Enhanced Wireless Network via Joint
  Active and Passive Beamforming},'' \emph{IEEE Transactions on Wireless
  Communications}, vol.~18, no.~11, pp. 5394--5409, 2019.

\bibitem{ma2019controllable}
Q.~Ma, L.~Chen, H.~B. Jing, Q.~R. Hong, H.~Y. Cui, Y.~Liu, L.~Li, and T.~J.
  Cui, ``{Controllable and Programmable Nonreciprocity based on Detachable
  Digital Coding Metasurface},'' \emph{Advanced Optical Materials}, vol.~7,
  no.~24, p. 1901285, 2019.

\bibitem{9520295}
W.~Jiang, B.~Chen, J.~Zhao, Z.~Xiong, and Z.~Ding, ``{Joint Active and Passive
  Beamforming Design for the IRS-Assisted MIMOME-OFDM Secure Communications},''
  \emph{IEEE Transactions on Vehicular Technology}, vol.~70, no.~10, pp.
  10\,369--10\,381, 2021.

\bibitem{9201173}
S.~Hong, C.~Pan, H.~Ren, K.~Wang, and A.~Nallanathan, ``{Artificial-Noise-Aided
  Secure MIMO Wireless Communications via Intelligent Reflecting Surface},''
  \emph{IEEE Transactions on Communications}, vol.~68, no.~12, pp. 7851--7866,
  2020.

\bibitem{9442833}
X.~Hu, L.~Jin, K.~Huang, X.~Sun, Y.~Zhou, and J.~Qu, ``{Intelligent Reflecting
  Surface-Assisted Secret Key Generation With Discrete Phase Shifts in Static
  Environment},'' \emph{IEEE Wireless Communications Letters}, vol.~10, no.~9,
  pp. 1867--1870, 2021.

\bibitem{9361290}
X.~Lu, J.~Lei, Y.~Shi, and W.~Li, ``{Intelligent Reflecting Surface Assisted
  Secret Key Generation},'' \emph{IEEE Signal Processing Letters}, vol.~28, pp.
  1036--1040, 2021.

\bibitem{9360860}
Z.~Ji, P.~L. Yeoh, G.~Chen, C.~Pan, Y.~Zhang, Z.~He, H.~Yin, and Y.~Li,
  ``{Random Shifting Intelligent Reflecting Surface for OTP Encrypted Data
  Transmission},'' \emph{IEEE Wireless Communications Letters}, vol.~10, no.~6,
  pp. 1192--1196, 2021.

\bibitem{staat2020intelligent}
P.~Staat, H.~Elders-Boll, M.~Heinrichs, R.~Kronberger, C.~Zenger, and C.~Paar,
  ``{Intelligent Reflecting Surface-Assisted Wireless Key Generation for
  Low-Entropy Environments},'' in \emph{2021 IEEE 32nd Annual International
  Symposium on Personal, Indoor and Mobile Radio Communications (PIMRC)}, 2021,
  pp. 745--751.

\bibitem{9298937}
Z.~Ji, P.~L. Yeoh, D.~Zhang, G.~Chen, Y.~Zhang, Z.~He, H.~Yin, and Y.~li,
  ``{Secret Key Generation for Intelligent Reflecting Surface Assisted Wireless
  Communication Networks},'' \emph{IEEE Transactions on Vehicular Technology},
  vol.~70, no.~1, pp. 1030--1034, 2021.

\bibitem{9625442}
L.~Hu, G.~Li, H.~Luo, and A.~Hu, ``{On the RIS Manipulating Attack and Its
  Countermeasures in Physical-layer Key Generation},'' in \emph{2021 IEEE 94th
  Vehicular Technology Conference (VTC2021-Fall)}, 2021, pp. 1--5.

\bibitem{9771319}
G.~Li, L.~Hu, P.~Staat, H.~Elders-Boll, C.~Zenger, C.~Paar, and A.~Hu,
  ``{Reconfigurable Intelligent Surface for Physical Layer Key Generation:
  Constructive or Destructive?}'' \emph{IEEE Wireless Communications}, vol.~29,
  no.~4, pp. 146--153, 2022.

\bibitem{thai2015physical}
C.~D.~T. Thai, J.~Lee, and T.~Q. Quek, ``{Physical-Layer Secret Key Generation
  with Colluding Untrusted Relays},'' \emph{IEEE Transactions on Wireless
  Communications}, vol.~15, no.~2, pp. 1517--1530, 2015.

\bibitem{letafati2020new}
M.~Letafati, A.~Kuhestani, D.~W.~K. Ng, and H.~Behroozi, ``{A New Frequency
  Hopping-Aided Secure Communication in the Presence of an Adversary Jammer and
  an Untrusted Relay},'' in \emph{2020 IEEE International Conference on
  Communications Workshops (ICC Workshops)}.\hskip 1em plus 0.5em minus
  0.4em\relax IEEE, 2020, pp. 1--7.

\bibitem{taha2021enabling}
A.~Taha, M.~Alrabeiah, and A.~Alkhateeb, ``{Enabling Large Intelligent Surfaces
  with Compressive Sensing and Deep Learning},'' \emph{IEEE Access}, vol.~9,
  pp. 44\,304--44\,321, 2021.

\bibitem{6847111}
A.~Alkhateeb, O.~El~Ayach, G.~Leus, and R.~W. Heath, ``{Channel Estimation and
  Hybrid Precoding for Millimeter Wave Cellular Systems},'' \emph{IEEE Journal
  of Selected Topics in Signal Processing}, vol.~8, no.~5, pp. 831--846, 2014.

\bibitem{9034493}
C.~You and R.~Zhang, ``{Hybrid Offline-Online Design for UAV-Enabled Data
  Harvesting in Probabilistic LoS Channels},'' \emph{IEEE Transactions on
  Wireless Communications}, vol.~19, no.~6, pp. 3753--3768, 2020.

\bibitem{9300189}
E.~Bj\"ornson and L.~Sanguinetti, ``{Rayleigh Fading Modeling and Channel
  Hardening for Reconfigurable Intelligent Surfaces},'' \emph{IEEE Wireless
  Communications Letters}, vol.~10, no.~4, pp. 830--834, 2021.

\bibitem{tsilipakos2020toward}
O.~Tsilipakos, A.~C. Tasolamprou, A.~Pitilakis, F.~Liu, X.~Wang, M.~S.
  Mirmoosa, D.~C. Tzarouchis, S.~Abadal, H.~Taghvaee, C.~Liaskos \emph{et~al.},
  ``{Toward Intelligent Metasurfaces: The Progress from Globally Tunable
  Metasurfaces to Software-Defined Metasurfaces with an Embedded Network of
  Controllers},'' \emph{Advanced optical materials}, vol.~8, no.~17, p.
  2000783, 2020.

\bibitem{8936989}
O.~\"Ozdogan, E.~Bj\"ornson, and E.~G. Larsson, ``{Intelligent Reflecting
  Surfaces: Physics, Propagation, and Pathloss Modeling},'' \emph{IEEE Wireless
  Communications Letters}, vol.~9, no.~5, pp. 581--585, 2020.

\bibitem{alexandropoulos2020hardware}
G.~C. Alexandropoulos and E.~Vlachos, ``{A Hardware Architecture for
  Reconfigurable Intelligent Surfaces with Minimal Active Elements for Explicit
  Channel Estimation},'' in \emph{ICASSP 2020-2020 IEEE International
  Conference on Acoustics, Speech and Signal Processing (ICASSP)}.\hskip 1em
  plus 0.5em minus 0.4em\relax IEEE, 2020, pp. 9175--9179.

\bibitem{candes2005decoding}
E.~J. Candes and T.~Tao, ``{Decoding by Linear Programming},'' \emph{IEEE
  Transactions on Information Theory}, vol.~51, no.~12, pp. 4203--4215, 2005.

\bibitem{candes2008restricted}
E.~J. Candes, ``{The Restricted Isometry Property and its Implications for
  Compressed Sensing},'' \emph{Comptes rendus mathematique}, vol. 346, no.
  9-10, pp. 589--592, 2008.

\bibitem{6006641}
T.~Zhang, ``{Sparse Recovery with Orthogonal Matching Pursuit under RIP},''
  \emph{IEEE Transactions on Information Theory}, vol.~57, no.~9, pp.
  6215--6221, 2011.

\bibitem{letafati2021deep}
M.~Letafati, H.~Behroozi, B.~H. Khalaj, and E.~A. Jorswieck, ``{Deep Learning
  for Hardware-Impaired Wireless Secret Key Generation with Man-in-the-Middle
  Attacks},'' in \emph{2021 IEEE Global Communications Conference
  (GLOBECOM)}.\hskip 1em plus 0.5em minus 0.4em\relax IEEE, 2021, pp. 1--6.

\bibitem{eberz2012practical}
S.~Eberz, M.~Strohmeier, M.~Wilhelm, and I.~Martinovic, ``{A Practical
  Man-in-the-Middle Attack on Signal-based Key Generation Protocols},'' in
  \emph{European symposium on research in computer security}.\hskip 1em plus
  0.5em minus 0.4em\relax Springer, 2012, pp. 235--252.

\bibitem{pan2021man}
Y.~Pan, Z.~Xu, M.~Li, and L.~Lazos, ``{Man-in-the-Middle Attack Resistant
  Secret Key Generation via Channel Randomization},'' in \emph{Proceedings of
  the Twenty-second International Symposium on Theory, Algorithmic Foundations,
  and Protocol Design for Mobile Networks and Mobile Computing}, 2021, pp.
  231--240.

\bibitem{8626511}
R.~Jin and K.~Zeng, ``{Manipulative Attack against Physical Layer Key Agreement
  and Countermeasure},'' \emph{IEEE Transactions on Dependable and Secure
  Computing}, vol.~18, no.~1, pp. 475--489, 2021.

\bibitem{8798662}
L.~Jin, S.~Zhang, Y.~Lou, X.~Xu, and Z.~Zhong, ``{Secret Key Generation With
  Cross Multiplication of Two-Way Random Signals},'' \emph{IEEE Access},
  vol.~7, pp. 113\,065--113\,080, 2019.

\bibitem{zhou2012pilot}
X.~Zhou, B.~Maham, and A.~Hjorungnes, ``{Pilot Contamination for Active
  Eavesdropping},'' \emph{IEEE Transactions on Wireless Communications},
  vol.~11, no.~3, pp. 903--907, 2012.

\bibitem{im2015secret}
S.~Im, H.~Jeon, J.~Choi, and J.~Ha, ``{Secret Key Agreement with Large Antenna
  Arrays under the Pilot Contamination Attack},'' \emph{IEEE Transactions on
  Wireless Communications}, vol.~14, no.~12, pp. 6579--6594, 2015.

\bibitem{billingsley1995probability}
P.~Billingsley, ``{Probability and Measure. 3rd Wiley},'' \emph{New York},
  1995.

\end{thebibliography}

\end{document}